\documentclass[twocolumn,aps,prb]{revtex4-1}
\usepackage{epsfig}
\usepackage{amsmath}
\usepackage{amssymb}
\usepackage[usenames]{color}
\usepackage{graphicx}
\usepackage{rotating}

\newcommand{\R}[1] {{\color{black} #1}}

\begin{document}

\date{\today}
\title{Feynman Diagram Description of 2D-Raman-THz Spectroscopy Applied to Water}
\author{David Sidler,Peter Hamm}
\affiliation{Department of Chemistry, University of Zurich, Zurich, Switzerland}
\date{\today}

\begin{abstract}
\textbf{Abstract:}
2D-Raman-THz spectroscopy of liquid water, which has been presented recently (Proc. Natl. Acad. Sci. USA 110, 20402 (2013)), directly probes the intermolecular degrees of freedom of the hydrogen-bond network. However, being a relatively new technique, its information content is not fully explored as to date. While the spectroscopic signal can be simulated based on molecular dynamics simulation in connection with a water force field, it is difficult to relate spectroscopic signatures to the underlying microscopic features of the force field. Here, a completely different approach is taken that starts from an as simple as possible model, i.e., a single vibrational mode with electrical and mechanical anharmonicity augmented with homogeneous and inhomogeneous broadening. An intuitive Feynman diagram picture is developed for all possible pulse sequences of hybrid 2D-Raman-THz spectroscopy. It is shown that the model can explain the experimental data essentially quantitatively with a very small set of parameters, \R{and it is tentatively concluded that the experimental signal originates from the hydrogen-bond stretching vibration around 170\,cm$^{-1}$. Furthermore, the echo observed in the experimental data can be quantified by fitting the model. A dominant fraction of its linewidth is attributed to quasi-inhomogeneous broadening in the slow-modulation limit with a correlation time of 370~fs, reflecting the lifetime of the hydrogen-bond networks giving rise the absorption band.}
\end{abstract}
\maketitle

\section{Introduction}

THz spectroscopy of liquids probes their intermolecular modes directly. Applied to liquid water, it reveals three distinct bands, which are assigned to hydrogen bond bending modes (50\,cm$^{-1}$), hydrogen bond vibrations (170\,cm$^{-1}$), and librations, which are hindered rotations of water molecules (600\,cm$^{-1}$).\cite{Bertie1996} These spectra can give us a picture of the motions in water, however, vibrational coherences are very short lived in water due to the fast, chaotic dynamics of the hydrogen bonding network, causing very broad and blurred bands. This limits the amount of information that can effectively be extracted from these spectra.

Extending the spectroscopy into two dimensions can thin out the information, and thus increase the amount of accessible information, and can furthermore disentangle homogeneous from inhomogeneous broadening.\R{\cite{tan93,okumura97a}} There has been significant effort to extend the very concept of  2D spectroscopy into the THz regime, either in the form of 2D Raman spectroscopy~\cite{VBout1991,Inaba1993,Muller1993,Tokmakoff1997,Blank1999,Blank2000,Kaufman2002,Golonzka2000,Kubarych2003,Li2008}, 2D-THz spectroscopy,\cite{Kuehn2009,Kuehn2011,Kuehn2011a,Elsaesser2015, Lu2016, Somma2016} or hybrid 2D-Raman-THz methods.\cite{Finneran2016,Finneran2017,Savolainen2013,Shalit2017}
Due to technical limitations, the latter is as of now the only 2D spectroscopy in the THz range that has been successfully applied to water and aqueous salt solutions.\cite{Savolainen2013,Shalit2017,Berger2018}
Of particular interest in these experiments is the observation of a very short-lived echo, whose decay time has been related to the relevant time-scales in water. However, a detailed understanding of the 2D-Raman-THz response is still lacking.

Unlike conventional 2D~IR spectroscopy,\cite{hamm11} \R{2D-Raman spectroscopy as well as 2D-Raman-THz spectroscopy} are described by second-order perturbation theory, despite the fact that it is a 3rd-order response with regard to the electrical fields. This is since the Raman process is electronically non-resonant and thus instantaneous, and it is assumed that the two field interactions giving rise to the Raman process occur at the same time. The three instead of four interactions of the system (including the emission process)  bring about that one has to induce a two-quantum transition, which would be forbidden in the harmonic approximation. This forbidden transition is a bottleneck of the signal, which is why its very cause (i.e., electrical or mechanical anharmonicity) determines the shape of the signal. Thus, the two-quantum transition must be an integral part of any model used to describe 2D-Raman-THz spectroscopy.

Time dependent perturbation theory is the starting point to calculate the 2D-Raman-THz response. The response functions for the three different time-orderings are:\cite{Cho1999}
\begin{align}
R_\mathrm{RTT} &= -\mathrm{Tr} \{ \hat\mu(t_1+t_2) [ \hat\mu(t_1), [ \hat\alpha(0), \rho_\mathrm{eq} ] ] \} \nonumber\\
R_\mathrm{TRT} &= -\mathrm{Tr} \{ \hat\mu(t_1+t_2) [ \hat\alpha(t_1), [ \hat\mu(0), \rho_\mathrm{eq} ] ] \} \nonumber\\
R_\mathrm{TTR} &= -\mathrm{Tr} \{ \hat\alpha(t_1+t_2) [ \hat\mu(t_1), [ \hat\mu(0), \rho_\mathrm{eq} ] ] \} \label{eq:response}
\end{align}
where $\hat\mu$ is the dipole operator, $\hat\alpha$ the polarizability operator, $t_1$ the time between the first and the second laser pulse interacting with the sample, and $t_2$  the time from the second laser pulse to the detection process. We have concentrated on the Raman-THz-THz (RTT) and the THz-Raman-THz (TRT) pulse sequences,\cite{Hamm2012,Hamm2012a,Savolainen2013,Hamm2014,Hamm2017,Shalit2017,Berger2018}
where the detection step measures an emitted THz field, while Blake and coworkers looked at the THz-THz-Raman (TTR) pulse sequence with a Raman process for  detection.\cite{Finneran2016, Finneran2017}

The theory of 2D-Raman-THz spectroscopy is similar to \R{2D-Raman spectroscopy,\cite{tan93,Steffen1996,Steffen1998,saito98,okumura97a,ma00,Jansen2000,Okumura2003,saito06,Hasegawa2006}} as well as hybrid IR-Raman techniques.\cite{Cho1999, Cho2000, Zhao2000, Guo2009,Grechko2018} In addition, a fair share of theory has been published tailored specifically for 2D-Raman-THz spectroscopy.\cite{Hamm2012, Hamm2012a, Hamm2014, Ito2014, Ito2015, Ikeda2015, Pan2015, Ito2016, Ito2016b}
Since typical THz experiments work in a frequency range equivalent to $k_BT$, the response can be derived in the classical limit from molecular dynamics (MD) simulation.\cite{Hasegawa2006} This approach appears to be the method of choice for complicated systems like water, since basically all effects, apart from possible quantum effects,\cite{Berger2018} are captured implicitly by a MD force field, including anharmonicities, mode coupling, chemical exchange, and orientational averaging. It has been shown that the MD approach reveals responses, which strongly depend on the force field used, especially on the description of polarizability, albeit in a rather nonintuitive and indirect way.\cite{Hamm2014, Ito2015} These MD simulations are largely ``black-box'' and it is difficult to disentangle the contributions to the 2D-Raman-THz response and to relate spectroscopic signatures to the underlying microscopic features of a water force field.

In order to learn more about the microscopic mechanism giving rise to the 2D-Raman-THz response, we take a completely different approach here and start from an as simple as possible model, i.e., a single vibrational mode augmented with homogeneous and inhomogeneous broadening. \R{To that end, we follow the conceptual framework introduced in  Ref.~\onlinecite{Okumura2003} for the description of 2D Raman spectroscopy, starting from a quantum-mechanical harmonic oscillator in an eigenstate representation and adding electrical and/or mechanical anharmonicity. The response functions Eq.~\ref{eq:response} can then} be depicted in a very intuitive way in terms of Feynman diagrams, from which one can directly read off the peak position in a 2D spectrum. We will see that the model can explain the experimental data essentially quantitatively with a small set of parameters.

\section{Model System}
\subsection{Zero-Order Description: Harmonic oscillator}
Due to its simplicity and the fact that it is often a very good approximation for molecular vibrations, the harmonic oscillator is an obvious starting point to set the stage:
\begin{equation}
\hat{H}^{(0)} = \frac{\hbar \omega}{2} \left(\hat p^2+\hat q^2\right).
\end{equation}
Analytical solutions of energy levels and eigenfunctions exist. From the thermal population of eigenstates, the equilibrium density matrix $\rho_\mathrm{eq}$ in Eq.~\ref{eq:response} can readily be constructed. Furthermore, the dimensionless position operator $\hat q = \sqrt{\frac{m\omega}{\hbar}} \hat x$  can be expressed in an harmonic oscillator eigenstate basis $| \varphi_i \rangle$
\begin{equation}
\left( \boldsymbol{q_H} \right) _{ij} \equiv \left \langle \varphi_i | \hat{q} | \varphi_j \right \rangle = \tfrac{1}{\sqrt{2}} \left [ \sqrt{j} \delta_{i+1,j} + \sqrt{i} \delta_{i-1,j} \right ] \label{eq:q}
\end{equation}
(where the subscript in $\boldsymbol{q_H}$ stands for ``harmonic''), which can be seen from the common creation ($\hat b$) and annihilation ($\hat b^\dag$) operator formalism with $\hat q=1/\sqrt{2}(\hat b^\dag+\hat b)$.\cite{Atkins2010} With that, the dipole and polarizability operators can be expanded in $\boldsymbol{q_H}$:
\begin{align}
\boldsymbol{\alpha} \propto \boldsymbol{q_H} + ... \nonumber\\
\boldsymbol{\mu} \propto \boldsymbol{q_H} + ...
\label{eq:linear}
\end{align}
The proportionality factors are irrelevant for the purpose of this paper, as they give rise to an overall intensity of the 2D-Raman-THz signal, which however is typically not determined experimentally in an absolute sense. The $\delta_{i+1,j}$ and $\delta_{i-1,j}$ terms in $\boldsymbol{q_H}$ couple states $i$ and $i\pm 1$, which leads to the well-known $i \rightarrow i\pm 1$ selection rules of the harmonic oscillator. This level of description is often sufficient to describe linear (1D) THz or Raman spectroscopy.

The $i \rightarrow i\pm 1$ selection rules cause coherence pathways, in which the density matrix is alternating between population and coherence states. As one starts from a thermal population state $\rho_\mathrm{eq}$, an even number of interactions would be needed to return to a population state after the emission of a signal. Therefore, due to the odd number of interactions in 2D-Raman-THz spectroscopy, the harmonic oscillator together with Eq.~\ref{eq:linear} would predict a vanishing signal. The appearance of a signal requires zero- or two-quantum transitions, which can be accomplished by breaking the symmetry of the system; either by considering nonlinearity of $\boldsymbol{\mu}$ and $\boldsymbol{\alpha}$ (electrical anharmonicity), or by perturbing the potential of the oscillator (mechanical anharmonicity).
Softening  the selection rules results in a bottleneck for the response, hence the signal shape is very sensitive to the specific nature of the anharmonicity, which thus is a crucial aspect of the modelling of  2D-Raman-THz spectra.

\subsection{Electrical Anharmonicity}

Electrical anharmonity is introduced by considering  higher order terms in equation Eq.~\eqref{eq:linear}:\R{\cite{Okumura2003}}
\begin{align}
\boldsymbol{\alpha} \propto \boldsymbol{q_H} + \sigma_\alpha \boldsymbol{q_H}^2+...\nonumber\\
\boldsymbol{\mu} \propto \boldsymbol{q_H} + \sigma_\mu \boldsymbol{q_H}^2+...
\label{eq:nonlinear}
\end{align}
with dimensionless smallness parameters  $|\sigma_\mu|\ll1$ and/or $|\sigma_\alpha|\ll1$. In an harmonic eigenstate basis,
the quadratic term is:
\begin{align}
(\boldsymbol{q_H}^2)_{ij} = \tfrac{1}{2} \Big[ (2i+1) \delta_{i,j} &+ \sqrt{(i+1)(i+2)} \delta_{i,j-2} + \nonumber\\
&\sqrt{(j+1)(j+2)}\delta_{i,j+2} \Big], \label{eq:elecAnharm}
\end{align}
which again  can be seen from the  creation/annihilation operator formalism.\cite{Atkins2010}
\R{In analogy to an electrical quadrupole transition}, the $\delta_{i,j}$-term allows for zero-quantum transitions $i \rightarrow i$, and the $\delta_{i,j\pm2}$-terms for two-quantum transitions $i \rightarrow i\pm 2$.

\subsection{Mechanical Anharmonicity}
Mechanical anharmonicity \R{(which has not been considered in Ref.~\onlinecite{Okumura2003})} breaks the symmetry by modifying the potential energy function. We consider a cubic anharmonic oscillator by adding $\sigma_\mathrm{M} \hbar \omega q^3$ to the harmonic Hamiltonian:
\begin{equation}
\hat{H} =  \hat{H}^{(0)} +  \sigma_\mathrm{M} \hbar \omega q^3
\end{equation}
with dimensionless smallness parameter $|\sigma_\mathrm{M}|\ll1$. The eigenstates of the anharmonic oscillator $|\varphi_i^{(anh)} \rangle$ can be calculated perturbatively.\cite{Claude1991} Although the position operator $\hat q$ \textit{per se} is not affected by this addition, its matrix representation in an anharmonic eigenstate basis is. It can be shown (see Appendix \ref{app:anharmonic}) that this matrix representation takes the form $\langle\varphi_i^{(anh)}|\hat q|\varphi_j^{(anh)}\rangle=(\boldsymbol{q_H})_{ij} + \sigma_\mathrm{M} (\boldsymbol{q_M})_{ij}$ with:
\begin{align}
\left(\boldsymbol{q_M} \right)_{ij} = \tfrac{1}{2} \Big[ - 3(2i + 1) \delta_{i,j} & + \sqrt{(i+1)(i+2)}\delta_{i,j-2} + \nonumber\\
& \sqrt{(j+1)(j+2)}\delta_{i,j+2} \Big], \label{eq:qM}
\end{align}
where the subscript ``M'' stands for ``mechanic anharmonicity''.
$\boldsymbol{q_M}$ has the same structure as $\boldsymbol{q_H}^2$ (Eq.~\ref{eq:elecAnharm}) with the $\delta_{i,j}$-term causing zero-quantum transitions and the $\delta_{i,j\pm2}$-terms causing two-quantum transitions, however the prefactors determining the transition probabilities and the signs of the peaks are different.

For a system with mechanical and electrical anharmonicity at the same time, the operators are:
\begin{align}
\boldsymbol{\alpha} &= \boldsymbol{q_H} + \sigma_\mathrm{M} \boldsymbol{q_M} + \sigma_\alpha \left( \boldsymbol{q_H} + \sigma_\mathrm{M} \boldsymbol{q_M} \right)^2 \nonumber\\
                    &\approx \boldsymbol{q_H} + \sigma_\mathrm{M} \boldsymbol{q_M} + \sigma_\alpha \boldsymbol{q_H}^2\nonumber\\
\boldsymbol{\mu} &= \boldsymbol{q_H} + \sigma_\mathrm{M} \boldsymbol{q_M} + \sigma_\mu \left(  \boldsymbol{q_H} + \sigma_\mathrm{M} \boldsymbol{q_M} \right)^2 \nonumber\\
                 &\approx \boldsymbol{q_H} + \sigma_\mathrm{M} \boldsymbol{q_M} + \sigma_\mu \boldsymbol{q_H}^2
\label{eq:dipolePolnonlinear}
\end{align}
where we neglected in the second step terms that are higher than first order in any of the smallness parameters $\sigma_\mu$, $\sigma_\alpha$ and $\sigma_\mathrm{M}$. For the same reason, we also neglect coherence pathways that contain more than one forbidden transition.

\begin{figure*}
\includegraphics[width=1\textwidth]{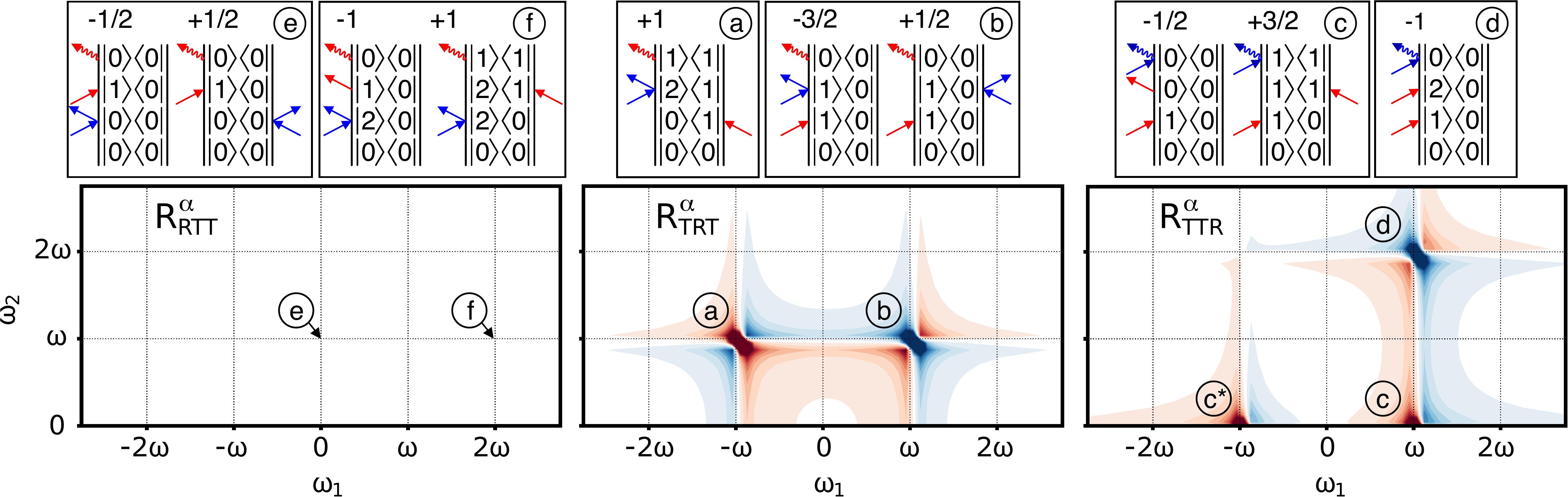}
\caption{Feynman diagrams and 2D frequency-domain spectra of $R_\mathrm{RTT}^\alpha$ (left), $R_\mathrm{TRT}^\alpha$ (middle) and $R_\mathrm{TTR}^\alpha$ (right), showing the real part.  The various 2D peaks are labeled with letters and the corresponding Feynman diagrams are shown above the spectra with THz interactions depicted in red and the Raman interaction as blue double-arrow. The weighting factor of each pathway is denoted above the Feynman diagram. The response functions were exponentially damped \R{according to Eq.~\ref{eq_vibRelax} with a time-constant long enough (i.e., $T_1=12\omega^{-1}$)} so that all peaks are resolved. Peak (c*) is the conjugate complex of peak (c).}
\label{fig:diag_raman}
\end{figure*}

\begin{figure*}
\includegraphics[width=1\textwidth]{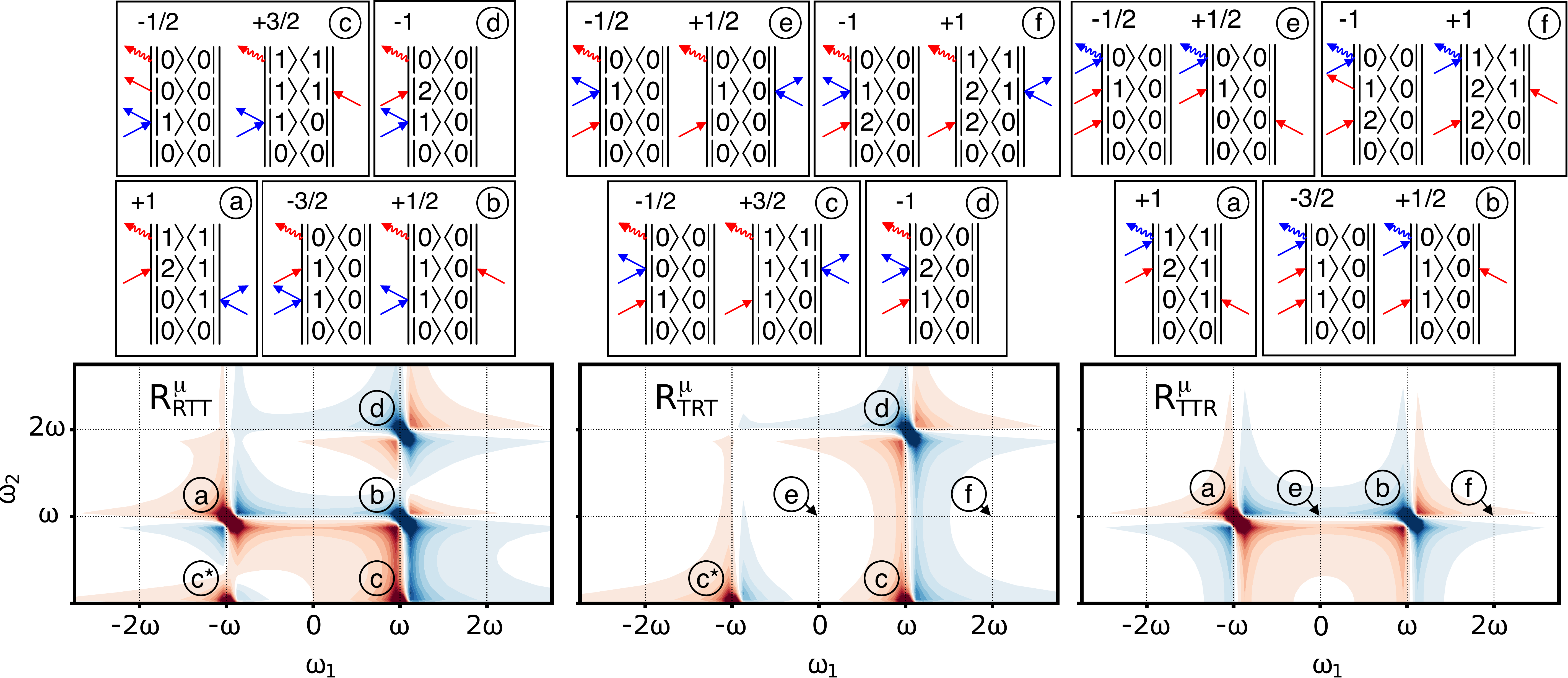}
\caption{Feynman diagrams and 2D frequency-domain  spectra of $R_\mathrm{RTT}^\mu$ (left), $R_\mathrm{TRT}^\mu$ (middle) and $R_\mathrm{TTR}^\mu$ (right); for a description see caption of Fig.~\ref{fig:diag_raman}. }
\label{fig:diag_thz}
\end{figure*}

\subsection{Feynman Diagrams}
\begin{figure}
\includegraphics[width=0.42\textwidth]{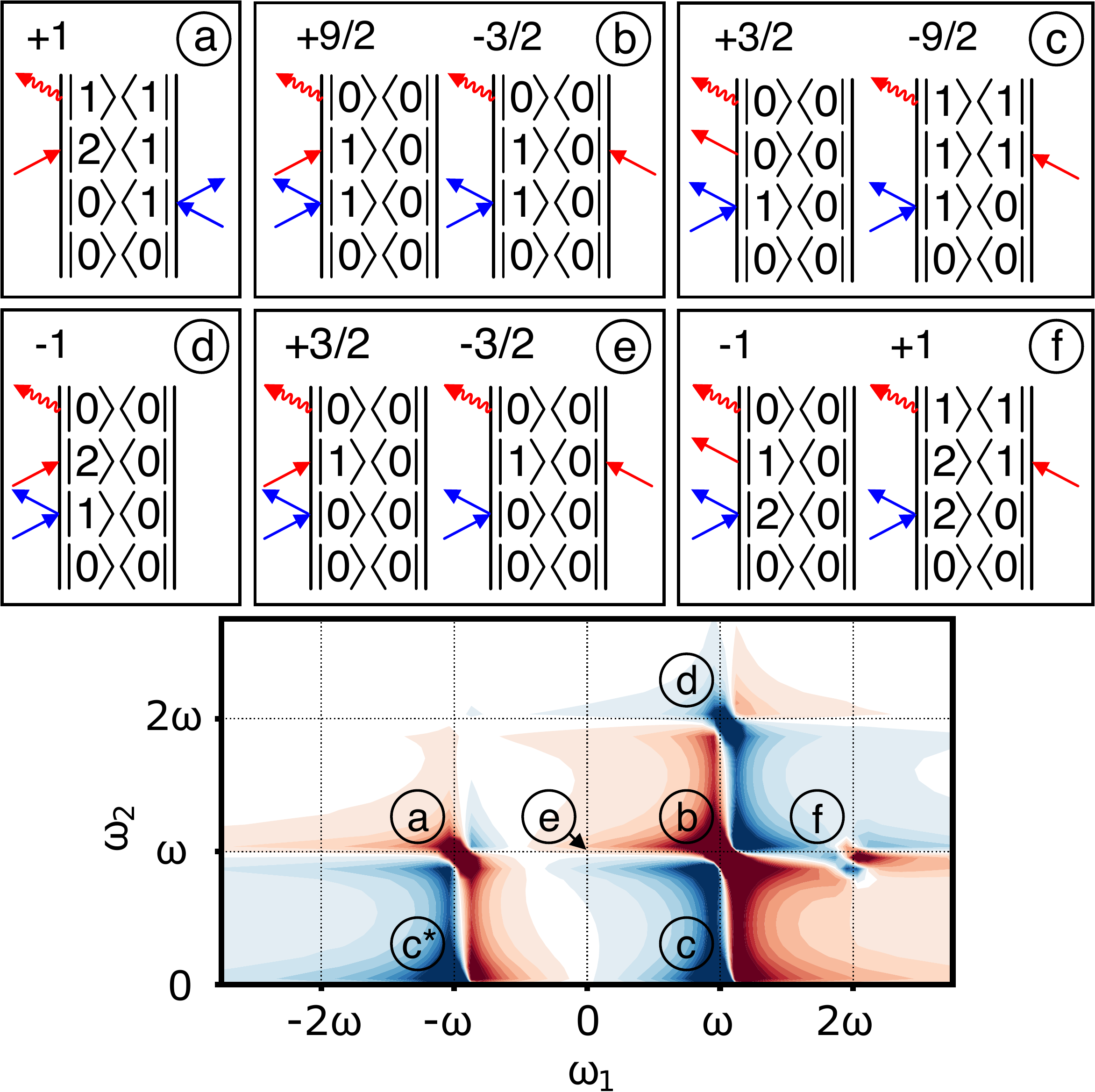}
\caption{Feynman diagrams and 2D frequency-domain  spectrum of $R_\mathrm{RTT}^M=R_\mathrm{TRT}^M=R_\mathrm{TTR}^M$; for a description see caption of Fig.~\ref{fig:diag_raman}.}
\label{fig:diag_mech}
\end{figure}

Based on  Eq.~\ref{eq:dipolePolnonlinear}, the response functions of Eq.~\ref{eq:response} can be separated into:
\begin{align}
R_\mathrm{RTT} \propto  \sigma_\alpha R_\mathrm{RTT}^{\alpha} +\sigma_\mu R_\mathrm{RTT}^\mu + \sigma_\mathrm{M} R_\mathrm{RTT}^\mathrm{M} \nonumber\\
R_\mathrm{TRT} \propto  \sigma_\alpha R_\mathrm{TRT}^{\alpha} +\sigma_\mu R_\mathrm{TRT}^\mu + \sigma_\mathrm{M} R_\mathrm{TRT}^\mathrm{M} \nonumber\\
R_\mathrm{TTR} \propto  \sigma_\alpha R_\mathrm{TTR}^{\alpha} +\sigma_\mu R_\mathrm{TTR}^\mu + \sigma_\mathrm{M} R_\mathrm{TTR}^\mathrm{M} \label{eq:Rsep}
\end{align}
with
\begin{align}
R_\mathrm{RTT}^\alpha = &-\mathrm{Tr}\left\{ \boldsymbol{q_H}(t_1+t_2) \left[ \boldsymbol{q_H}(t_1), \left[ \boldsymbol{q_H}^2(0) , \rho_\mathrm{eq} \right] \right] \right\} \nonumber\\
R_\mathrm{TRT}^\alpha = &-\mathrm{Tr}\left\{ \boldsymbol{q_H}(t_1+t_2) \left[ \boldsymbol{q_H}^2(t_1), \left[ \boldsymbol{q_H}(0) , \rho_\mathrm{eq} \right] \right] \right\} \nonumber\\
R_\mathrm{TTR}^\alpha = &-\mathrm{Tr}\left\{ \boldsymbol{q_H}^2(t_1+t_2) \left[ \boldsymbol{q_H}(t_1), \left[ \boldsymbol{q_H}(0) , \rho_\mathrm{eq} \right] \right] \right\} \nonumber\\
R_\mathrm{RTT}^\mu = &-\mathrm{Tr}\left\{ \boldsymbol{q_H}(t_1+t_2) \left[ \boldsymbol{q_H}^2(t_1), \left[ \boldsymbol{q_H}(0) , \rho_\mathrm{eq} \right] \right] \right\} \nonumber\\
 &- \mathrm{Tr} \left\{ \boldsymbol{q_H}^2(t_1+t_2) \left[ \boldsymbol{q_H}(t_1) , \left[ \boldsymbol{q_H}(0) , \rho_\mathrm{eq} \right] \right] \right\} \nonumber \\
R_\mathrm{TRT}^\mu = &-\mathrm{Tr}\left\{ \boldsymbol{q_H}(t_1+t_2) \left[ \boldsymbol{q_H}(t_1), \left[ \boldsymbol{q_H}^2(0) , \rho_\mathrm{eq} \right] \right] \right\} \nonumber\\
 & - \mathrm{Tr} \left\{ \boldsymbol{q_H}^2(t_1+t_2) \left[ \boldsymbol{q_H}(t_1) , \left[ \boldsymbol{q_H}(0) , \rho_\mathrm{eq} \right] \right] \right\} \nonumber \\
R_\mathrm{TTR}^\mu = &-\mathrm{Tr}\left\{ \boldsymbol{q_H}(t_1+t_2) \left[ \boldsymbol{q_H}(t_1), \left[ \boldsymbol{q_H}^2(0) , \rho_\mathrm{eq} \right] \right] \right\} \nonumber\\
 & - \mathrm{Tr} \left\{ \boldsymbol{q_H}(t_1+t_2) \left[ \boldsymbol{q_H}^2(t_1) , \left[ \boldsymbol{q_H}(0) , \rho_\mathrm{eq} \right] \right] \right\} \nonumber \\
R_\mathrm{RTT}^\mathrm{M} = &-\mathrm{Tr}\left\{ \boldsymbol{q_M}(t_1+t_2) \left[ \boldsymbol{q_H}(t_1), \left[ \boldsymbol{q_H}(0) , \rho_\mathrm{eq} \right] \right] \right\} \nonumber\\
 &- \mathrm{Tr} \left\{ \boldsymbol{q_H}(t_1+t_2) \left[ \boldsymbol{q_M}(t_1) , \left[ \boldsymbol{q_H}(0) , \rho_\mathrm{eq} \right] \right] \right\} \nonumber \\
 &- \mathrm{Tr} \left\{ \boldsymbol{q_H}(t_1+t_2) \left[ \boldsymbol{q_H}(t_1) , \left[ \boldsymbol{q_M}(0) , \rho_\mathrm{eq} \right] \right] \right\} \nonumber \\
 &=  R_\mathrm{TRT}^\mathrm{M} =  R_\mathrm{TTR}^\mathrm{M}, \label{eq:responseAll}
\end{align}
which allows us to discuss the contributions from electrical ($\sigma_\alpha$,$\sigma_\mu$) and mechanical anharmonicity ($\sigma_\mathrm{M}$) separately. To this end, we expand the nested commutators in Eqs.~\ref{eq:responseAll} and represent all terms in the form of Feynman diagrams (Figs.~\ref{fig:diag_raman}, \ref{fig:diag_thz} and \ref{fig:diag_mech}). The corresponding 2D peaks are also summarized in Tab.~\ref{tab:intensities} together with their intensities, which originate from products of the prefactors of the corresponding $\delta$-terms in Eqs.~\ref{eq:q}, \ref{eq:elecAnharm} and \ref{eq:qM}, and their sums when contributions overlap in frequency space. \R{The frequency positions of the in total four peaks, ($-\omega, \omega$), ($\omega, \omega$), ($\omega,0$), and ($\omega, 2\omega$), are the same as in Ref.~\onlinecite{Okumura2003}, but their intensities differ since the selection rules of the three pulse sequences of 2D-Raman-THz spectroscopy differ from those in 2D-Raman spectroscopy.} The various spectra are discussed in the following.

\begin{table}[b]
\centering
\caption{Intensities of all non-zero peaks appearing in Eq.~\eqref{eq:responseAll}. }
\label{tab:intensities}
\begin{tabular}{c |  c c c c}
 & \multicolumn{4}{c}{Peak } \\
 & (a) & (b) & (c) & (d) \\
	Response & ($-\omega, \omega$) & ($\omega, \omega$) & ($\omega,0$) & ($\omega, 2\omega$) \\ \hline
$R_\mathrm{RTT}^\alpha$ & 0 & 0 & 0 & 0 \\
$R_\mathrm{TRT}^\alpha$ & 1 & -1 & 0 & 0 \\
$R_\mathrm{TTR}^\alpha$ & 0 & 0& 1 & -1 \\
$R_\mathrm{RTT}^\mu$ & 1 & -1 & 1 & -1  \\
$R_\mathrm{TRT}^\mu$ & 0 & 0 & 1 & -1 \\
$R_\mathrm{TTR}^\mu$ & 1 & -1 & 0 & 0 \\
$R_\mathrm{RTT}^M$ & 1 & 3  & -3 & -1 \\
$R_\mathrm{TRT}^M$ & 1 & 3 & -3 & -1 \\
$R_\mathrm{TTR}^M$ & 1 & 3 & -3 & -1
\end{tabular}
\end{table}

In $R_\mathrm{RTT}^\alpha$, $R_\mathrm{TRT}^\alpha$, and $R_\mathrm{TTR}^\alpha$ (Fig.~\ref{fig:diag_raman}), the nonlinearity of $\boldsymbol{\alpha}$ allows the Raman interaction to induce a zero- or a two-quantum transition via the contribution of $\boldsymbol{q_H}^2$.
Starting with $R_\mathrm{RTT}^\alpha$, four Feynman diagrams can be constructed (Fig.~\ref{fig:diag_raman}, top left). In this case the first Raman interaction  induces a forbidden transition, i.e. a zero-quantum or a two-quantum transition. The pathways contributing to peak (e) start with a zero-quantum transition from either the left and the right, which however lead to two equivalent contributions of opposite sign that perfectly cancel out. Peak (f), on the other hand, starts with a two-quantum transition from the left. From there, the system goes into a $|1\rangle \langle 0|$ coherence by the second interaction from the left, or into $|2 \rangle \langle 1|$ coherence via an interaction from the right. For a harmonic potential, the energy levels are equally spaced and both coherences oscillate with the same frequency. \R{If we furthermore assume that the 0-1 and the 1-2 dephasing times are the same (see Eq.~\ref{eq_vibRelax} below),\cite{Okumura2003}} both contributions again cancel, and we have $R_\mathrm{RTT}^\alpha = 0$ altogether.

In contrast, two peaks appear in the TRT pulse sequence. For peak (a), the first interaction brings the system into a $|0\rangle\langle 1|$ coherence, and the second Raman interaction does a two-quantum transition to bring the system into the $|2\rangle \langle 1 |$ coherence, hence the peak appears  at $(-\omega, \omega)$. Peak (b), in contrast, is diagonal at $(\omega, \omega)$, since  the second interaction does a zero-quantum transition, which does not affect the state. The zero-quantum transition can either act from the left or the right side, giving rise to two contributions with opposite sign. However, these contributions do not fully cancel, since the interaction from left is weighted by the $(\boldsymbol{q_H}^2)_{1,1} = -3/2$ element, and the interaction from right by $(\boldsymbol{q_H}^2)_{0,0} = +1/2$. Finally, the TTR sequence has its forbidden transition in the emission step with zero-quantum coherences for peak (c), and two-quantum coherences for peak (d). Peak (c), with $\omega_2=0$, appears together with its conjugate complex (c$^*$). The conjugate complex of all other peaks show up at negative frequencies $\omega_2$, but the negative $\omega_2$-half-space is not shown in Fig.~\ref{fig:diag_raman} for clarity.

There are more peaks in $R_\mathrm{RTT}^\mu$, $R_\mathrm{TRT}^\mu$, and $R_\mathrm{TTR}^\mu$ (Fig.~\ref{fig:diag_thz}), since there are two interactions with $\boldsymbol{\mu}$ and the forbidden transition can occur with either one of it. However, Eq.~\ref{eq:responseAll} shows that each term of $R^\mu$ can be written as a sum of two terms that we already discussed in the context of $R^\alpha$, e.g.
$R_\mathrm{RTT}^\mu = R_\mathrm{TRT}^\alpha + R_\mathrm{TTR}^\alpha$. The corresponding 2D spectra in Fig.~\ref{fig:diag_thz} therefore are overlays of two spectra each from Fig.~\ref{fig:diag_raman}.

Finally, the responses $R_\mathrm{RTT}^\mathrm{M}$, $R_\mathrm{TRT}^\mathrm{M}$ and $R_\mathrm{TTR}^\mathrm{M}$ originating from mechanical anharmonicity are shown  in Fig.~\ref{fig:diag_mech}. The term $\sigma_\mathrm{M} \boldsymbol{q_M}$ contributes to $\boldsymbol{\mu}$ and $\boldsymbol{\alpha}$ equally, and the forbidden transition can occur at any of the tree interactions. Since $\boldsymbol{\mu} = \boldsymbol{\alpha}$ for this case, the order of interactions does not matter, and $R_\mathrm{RTT}^\mathrm{M} = R_\mathrm{TRT}^\mathrm{M}= R_\mathrm{TTR}^\mathrm{M}$. The corresponding 2D spectrum is in essence an overlay of all three spectra shown in  Fig.~\ref{fig:diag_raman}, except for the fact that the features that include zero-quantum transitions (b,c) change signs and amplitudes due to the different diagonal element of $\boldsymbol{q_M}$. Furthermore, peak (f) is now nonzero due to the anharmonic shift of the energy levels.

\R{As in Ref.~\onlinecite{Okumura2003},} we considered so far only Feynman diagrams starting from the $|0\rangle\langle0|$ element of the density matrix in Figs.~\ref{fig:diag_raman}, \ref{fig:diag_thz} and \ref{fig:diag_mech}, implicitly assuming a temperature $T$=0~K. It can however be shown that the overall response function does not depend on the starting level, despite the fact that more Feynman diagrams come into play, and hence is in fact temperature independent (see Appendix \ref{sec:temperature} for details). For that we however need to assume \R{that the lineshape functions depend on level differences only (see Eq.~\ref{eq_vibRelax} below)}, and that the effect of mechanical anharmonicity is on the transitions probabilities only (via $\boldsymbol{q_M}$), while we keep the energy spectrum equidistant. Fig.~\ref{fig:diag_mech} shows that this is a good approximation. That is,  the anharmonic shift $\omega_{21}-\omega_{10}$ is small, and for any reasonable, not too narrow spectral linewidth both contributions giving rise to peak (f) largely cancel each other. In comparison, the effect of mechanical anharmonicity on the transition probabilities, giving rise to peaks (a)-(e), is much larger. Within that approximation, quantum and classical response functions are in fact identical, and \R{simple analytic expressions can be derived} for the three response functions by collecting all terms \R{(without that approximation, an explicit sum over Boltzmann-weighted initial states would be needed that converges extremely slowly for $\hbar\omega\ll k_BT$)}:
\R{
\begin{widetext}
\begin{align}
R_\mathrm{RTT}(t_1,t_2)\propto\Theta(t_1,t_2)\big(&(\sigma_\mu + \sigma_\mathrm{M}) \cos(\omega t_1 - \omega t_2)\Gamma_{-\omega,\omega}(t_1,t_2) + (-\sigma_\mu + 3 \sigma_\mathrm{M})  \cos(\omega t_1 + \omega t_2)\Gamma_{\omega,\omega}(t_1,t_2) \nonumber \\
+ &  (\sigma_\mu - 3 \sigma_\mathrm{M}) \cos(\omega t_1)\Gamma_{\omega,0}(t_1,t_2)+ (-\sigma_\mu - \sigma_\mathrm{M}) \cos(\omega t_1 + 2 \omega t_2)\Gamma_{\omega,2\omega}(t_1,t_2) \big)  \nonumber \\
R_\mathrm{TRT}(t_1,t_2) \propto\Theta(t_1,t_2)\big(& (\sigma_\alpha + \sigma_\mathrm{M}) \cos(\omega t_1 - \omega t_2) \Gamma_{-\omega,\omega}(t_1,t_2) + (-\sigma_\alpha + 3 \sigma_\mathrm{M}) \cos(\omega t_1 + \omega t_2) \Gamma_{\omega,\omega}(t_1,t_2)\nonumber\\
+& (\sigma_\mu - 3 \sigma_\mathrm{M}) \cos(\omega t_1)\Gamma_{\omega,0}(t_1,t_2) + (-\sigma_\mu - \sigma_\mathrm{M}) \cos(\omega t_1 + 2 \omega t_2)\Gamma_{\omega,2\omega}(t_1,t_2) \big)   \nonumber\\
R_\mathrm{TTR}(t_1,t_2)\propto\Theta(t_1,t_2)\big(& (\sigma_\mu + \sigma_\mathrm{M}) \cos(\omega t_1 - \omega t_2) \Gamma_{-\omega,\omega}(t_1,t_2) + (-\sigma_\mu + 3 \sigma_\mathrm{M}) \cos(\omega t_1 + \omega t_2)\Gamma_{\omega,\omega}(t_1,t_2)  \nonumber\\
+&  (\sigma_\alpha - 3 \sigma_\mathrm{M}) \cos(\omega t_1) \Gamma_{\omega,0}(t_1,t_2) + (-\sigma_\alpha - \sigma_\mathrm{M}) \cos(\omega t_1 + 2 \omega t_2)\Gamma_{\omega,2\omega}(t_1,t_2) \big) \label{eq:Analytic}
\end{align}
\end{widetext}
}
where $\Theta(t_1,t_2)$ is the Heaviside function ensuring that the response vanishes for $t_1<0$ or $t_2<0$. In second order perturbation theory, the complex conjugate of a term can be generated by two permutations in the commutator. Therefore, the two complex conjugate components have the same sign and reveal $\cos$-functions upon addition.

\R{We augment these response functions with dephasing terms $\Gamma(t_1,t_2)$ in essence along the lines of Ref.~\onlinecite{Steffen1998}, but replacing the part originating from vibrational $T_1$ relaxation with Eq.~36, rather than Eq.~37, from Ref.~\onlinecite{Okumura2003}.  That is, each response function in Eq.~\ref{eq:Analytic}
contains one term each for peaks at $(-\omega,\omega)$,  $(\omega,\omega)$, $(\omega,0)$ and $(\omega,2\omega)$, respectively, which are multiplied with the following damping functions:
\begin{align}
\Gamma_{-\omega,\omega}(t_1,t_2)=&e^{-\frac{t_1+t_2}{2T_1}}e^{-2 g(t_1)-2 g(t_2) +g(t_1+t_2)} \nonumber\\
\Gamma_{\omega,\omega}(t_1,t_2)=&e^{-\frac{t_1+t_2}{2T_1}}e^{-g(t_1+t_2)} \nonumber\\
\Gamma_{\omega,0}(t_1,t_2)=&e^{-\frac{t_1+2t_2}{2T_1}}e^{-g(t_1)} \nonumber\\
\Gamma_{\omega,2\omega}(t_1,t_2)=&e^{-\frac{t_1+2t_2}{2T_1}}e^{+g(t_1)-2g(t_2)-2g(t_1+t_2)}
\end{align}
As discussed in Ref.~\onlinecite{Okumura2003}, the damping terms caused by vibrational relaxation depend on quantum numbers according to:
\begin{align}
T_1^{(n,n)}&=T_1 \nonumber\\
T_2^{(n,m)}&=\frac{2T_1}{|n-m|}. \label{eq_vibRelax}
\end{align}
The lineshape function $g(t)$ is:
\begin{equation}
g(t)=\Delta\omega^2\tau_c^2\left(e^{-t/\tau_c}+t/\tau_c-1\right), \label{eq_gt}
\end{equation}
whose origin are Gaussian fluctuations of the transition frequency $\omega(t)$ with standard deviation $\Delta\omega$ and correlation time $\tau_c$:
\begin{equation}
\langle\delta\omega(t)\delta\omega(0)\rangle=\Delta\omega^2e^{-t/\tau_c},
\end{equation}
where $\delta\omega(t)=\omega(t)-\langle\omega(t)\rangle$. In the limes $\Delta\omega\tau_c\ll1$, Eq.~\ref{eq_gt} causes pure dephasing with $T^*_2=(\Delta\omega^2\tau_c)^{-1}$, in the limes $\Delta\omega\tau_c\gg1$ inhomogeneous dephasing with a Gaussian lineshape with width $\Delta\omega$.}

\subsection{Inhomogeneous Broadening and Echoes}

\begin{figure}
\includegraphics[width=0.45\textwidth]{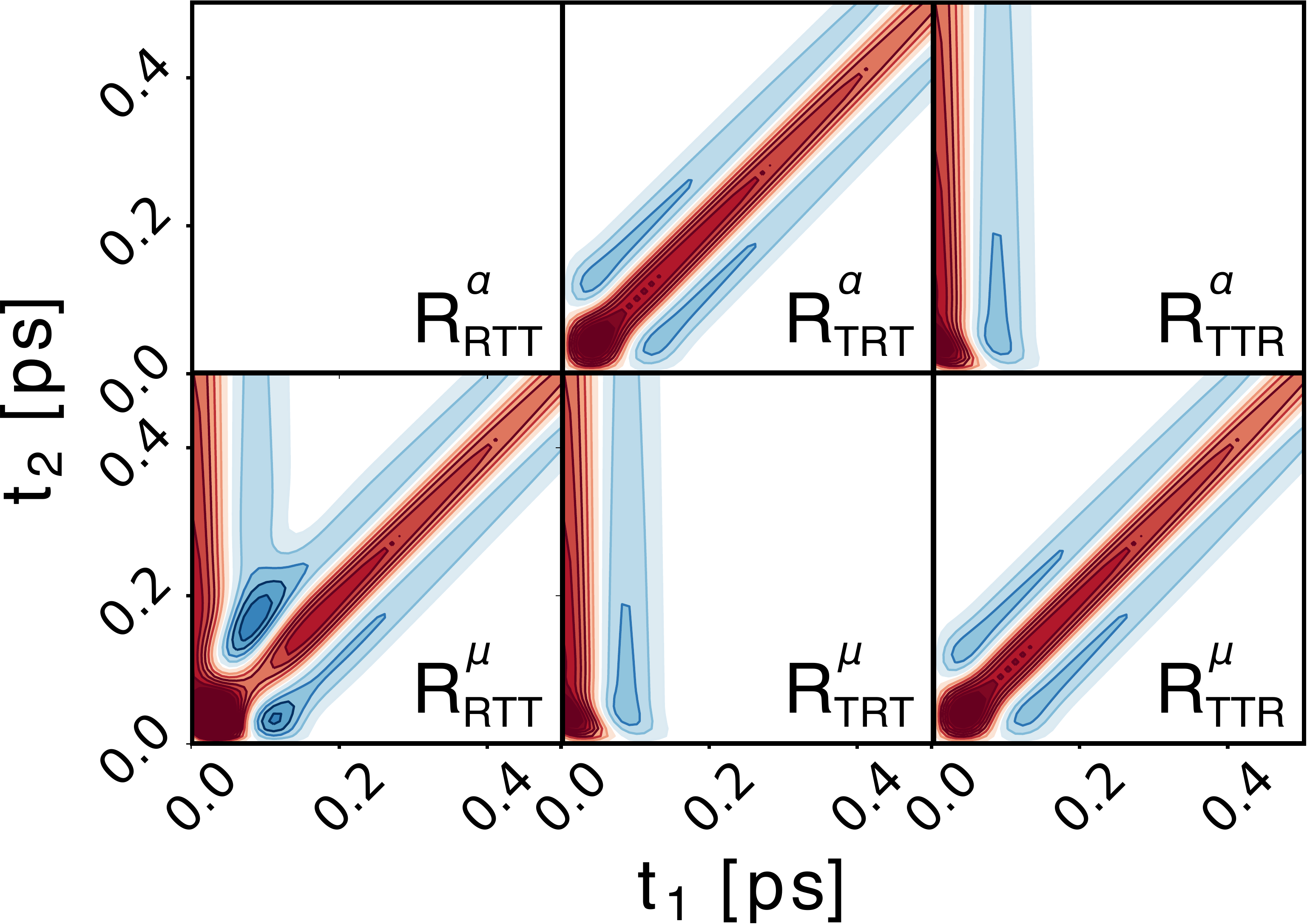}
\caption{Time-domain signals $R_\mathrm{RTT}$, $R_\mathrm{TRT}$, and $R_\mathrm{TTR}$ for an inhomogeneously broadened system, \R{using the parameters from Tab.~\ref{tab:fit} for the hydrogen-bond stretching vibration, but setting $\tau_c=$100~ps and $T_1=1~$ps.} The effects of nonlinear polarizability is shown in the top row, and that of nonlinear dipole moment in the bottom row.}
\label{fig:echoes}
\end{figure}

In any of the responses, peak (a) with frequencies of opposite signs $(-\omega,\omega)$ is a rephasing pathway,\cite{Hamm2017} as a coherence that dephased during the first time period might rephase during the second, provided the band is inhomogeneously broadened and maintains some amount of memory on the oscillation frequency. Rephasing requires an ``inversion of coherence'', and the only possibility to achieve that
are coherence pathway that start with a $|0\rangle\langle 1|$ coherence and are then brought into a $|2\rangle \langle 1 |$ coherence by the second pulse (see Figs.~\ref{fig:diag_raman}, \ref{fig:diag_thz} and \ref{fig:diag_mech}). In a time-domain representation, rephasing will generate an echo along $t_1=t_2$; in a frequency-domain representation tilted 2D lineshapes. Peak (b), in contrast, is a non-rephasing pathway.

To explore the appearance of echoes (Fig.~\ref{fig:echoes}), \R{we choose lineshape parameters for the  hydrogen-bond stretching vibration as listed in Tab.~\ref{tab:fit} (see below), except of unrealistically long values for $\tau_c=$100~ps and $T_1=1~$ps.} Echoes along $t_1=t_2$ are seen for $R_{TRT}^\alpha$, $R_{RTT}^\mu$, and $R_{TTR}^\mu$, i.e., coherence pathways with their two-quantum transition for the second pulse to induce the required two-quantum transition. Experimentally, one can distinguish between $R_{RTT}$, $R_{TRT}$, and $R_{TTR}$ by the choice of a pulse sequence, hence when observing an echo, one gets an handle on determining the major source of anharmonicity.  The vertical features in $R_{TTR}^\alpha$, $R_{RTT}^\mu$ and $R_{TRT}^\mu$ originate from peak (c), which does not experience any inhomogeneous dephasing in the $t_2$ direction, since its frequency in $\omega_2$ is always zero.

\section{Comparison to Water Experiments}

\subsection{Instrument Response Function}
We now explore to what extent the model derived here can explain the experimental data of liquid water from Ref.~\onlinecite{Savolainen2013}. We start with noting that this experiment measured the $R_\mathrm{RTT}$ and $R_\mathrm{TRT}$ responses by interchanging the timing between the Raman and THz-pump pulses, but not the $R_\mathrm{TTR}$ response, which would require a different detection scheme.\cite{Finneran2016, Finneran2017} With the particular arrangement of the delay lines in the experiment of Ref.~\onlinecite{Savolainen2013}, we measured:
\begin{equation}
R(t_1, t_2) = R_\mathrm{RTT}(t_1, t_2) + R_\mathrm{TRT}(-t_1, t_2-t_1), \label{eq:paste}
\end{equation}
where time $t_1$ is from the Raman pump pulse to the THz pump pulse and time $t_2$ from the THz pump pulse to  detection (which is why $t_2-t_1$ appears as argument of $R_\mathrm{TRT}$ where the THz pump comes before the Raman pump, while $t_2$ is the time from the second Raman interaction to detection in the definition of $R_\mathrm{TRT}$ in Eq.~\ref{eq:Analytic}). Furthermore, the experimental signal is obtained from the response functions by a convolution with the instrument response function (IRF):
\begin{equation}
S(t_1, t_2) =	I(t_1, t_2) \circledast R(t_1,t_2), \label{eq:convolution}
\end{equation}
which is calculated from the THz field $E_\mathrm{THz}$ and the envelope of the Raman pulse $I_\mathrm{Raman}$:
\begin{equation}
I(t_1,t_2) \propto \frac{d}{dt_2} E_\mathrm{THz}(t_2) I_\mathrm{Raman}(t_2 + t_1). \label{eq:IRF}
\end{equation}
Eq.~\ref{eq:IRF} is an idealized expression to illustrate the basic idea; in the real experiment the IRF contains in addition a transfer function describing how the emitted THz field reshapes as it propagates from the sample to the detection crystal, as well as a correction for a non-perfect Gouy phase (see Refs.~\onlinecite{Savolainen2013,Ahmed2014} for details).

\begin{figure}
\includegraphics[width=0.35\textwidth]{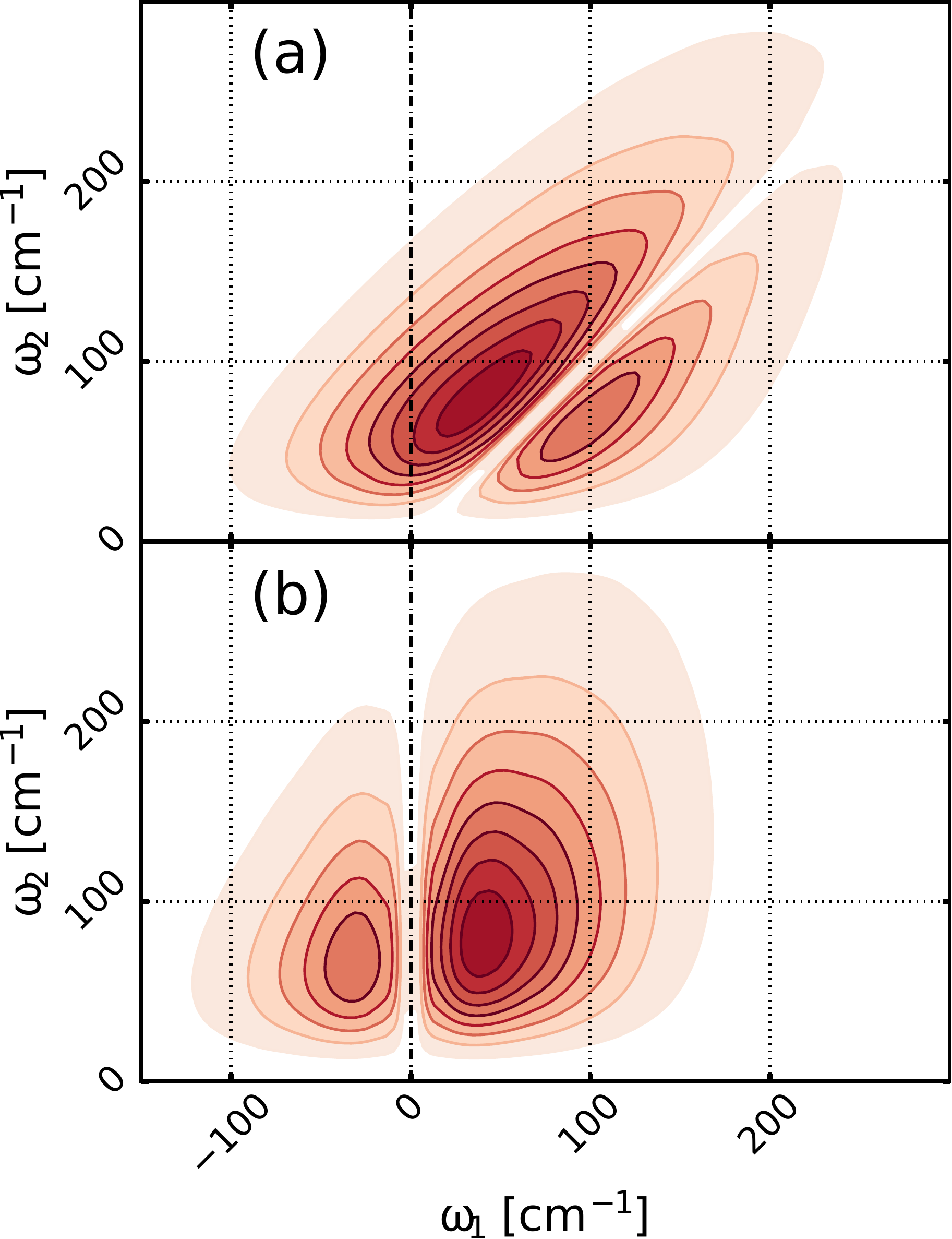}
\caption{The absolute value of the IRF in the frequency domain for (a) the RTT pulse sequence and (b) the TRT pulse sequence, which have been derived from Eq.~\ref{eq:IRF} with the experimentally determined pulses and transfer function.\cite{Savolainen2013}}
\label{fig:irf}
\end{figure}

It is illustrative to look at the IRF in the frequency domain (Fig.~\ref{fig:irf}), as it shows which one of the peaks in the 2D responses (Figs.~\ref{fig:diag_raman}, \ref{fig:diag_thz} and \ref{fig:diag_mech}) the experiment is sensitive to. The IRF spectrum  for the RTT pulse sequence (Fig. \ref{fig:irf}a) has two nodal lines, one at $\omega_2 = 0$, which is caused by the derivative $d/dt_2$ in Eq.~\ref{eq:IRF}, and one at $\omega_1 = \omega_2$, which caused by the fact that the THz pump pulse does not have any DC component (i.e., $\int E_\mathrm{THz}(t) dt = 0$). Peaks (b) and (c) lie on these nodal lines, as they would contain zero-quantum transitions for either the THz pump pulse or the THz emission process. We will nevertheless see these peaks to a certain extent since the damping in water is very fast, hence they have a significant spectral width that extends into regions where the IRF is non-zero.  Furthermore, a smaller bandwidth is observed in the rephasing quadrant ($\omega_1 < 0$, $\omega_2 > 0$) than in the non-rephasing quadrant ($\omega_1, \omega_2 > 0$),  reflecting the fact that the peak at $\omega_2 = -\omega_1$ includes a two-quantum transition which requires a higher bandwidth of the THz pulse.

For the $R_\mathrm{TRT}$ response, the IRF has to be transformed to account for the time shift applied in Eq.~\ref{eq:paste} (Fig. \ref{fig:irf}b).
The effect of a time transform $t_1 = -t_1'$ and $t_2 = t_2' - t_1'$ on the frequency domain can be derived by inspecting the effect on a single basis function of the Fourier transformation:
\begin{align}
\exp \left( i \left[ \omega_1 t_1 + \omega_2 t_2  \right] \right)  &= \exp \left( i \left[ \omega_1 (-t_1' ) + \omega_2 (t_2' - t_1') \right] \right) \nonumber \\
&= \exp \left( i \left[ \omega_1' t_1' + \omega_2' t_2' \right] \right) \
\end{align}
with $\omega_1' = -\omega_1 - \omega_2$ and $\omega_2' = \omega_2$.
The transformed instrument response has nodal lines at $\omega_1 = 0$ and $\omega_2 = 0$. Hence, in this case, the diagonal peak (b) lies in a region where the IRF is large, since the envelope of the Raman pulse has a zero-frequency component.

\subsection{1D Spectra and 2D Response}

\begin{figure}
\includegraphics[width=0.35\textwidth]{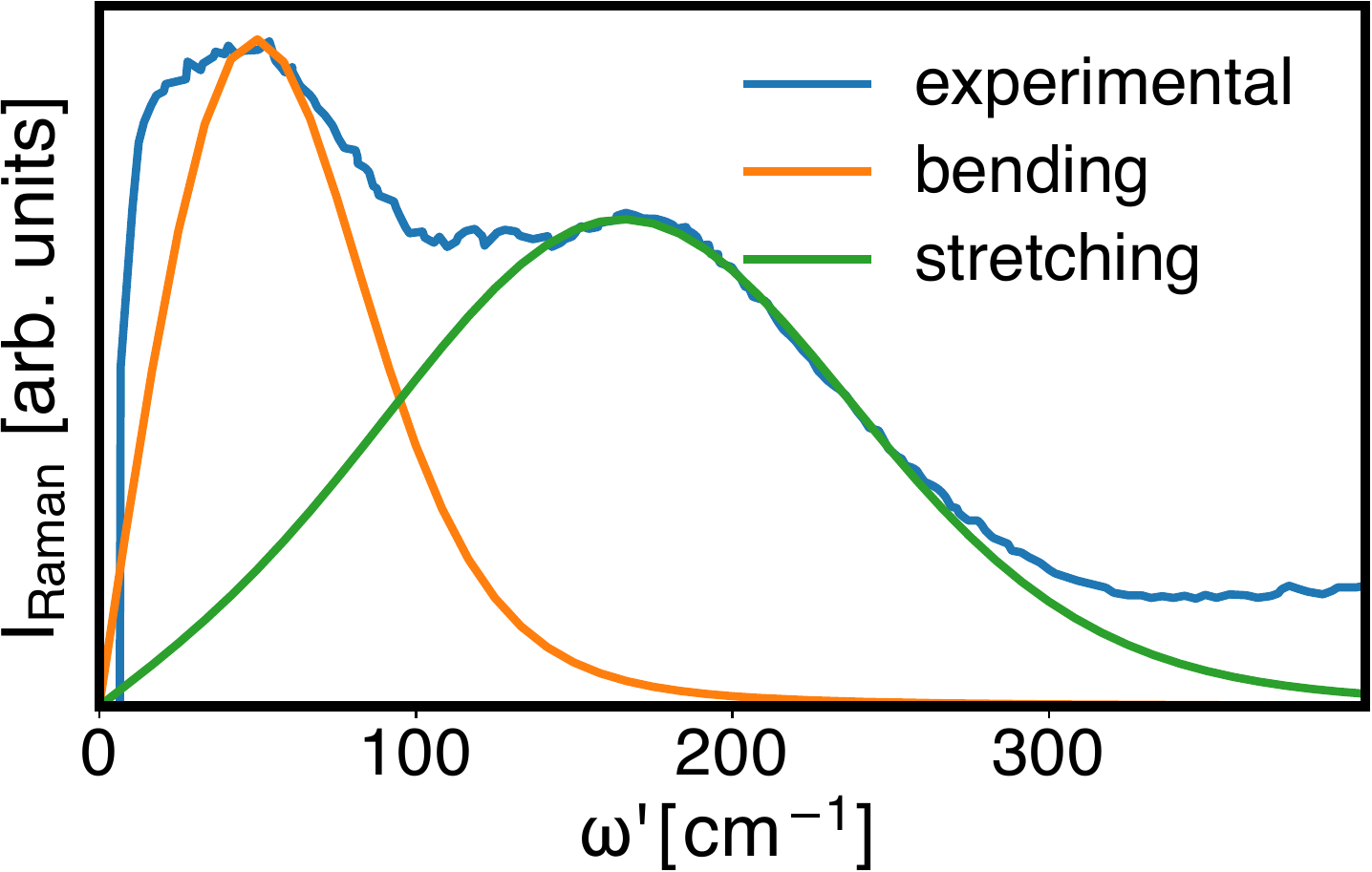}
\caption{The experimental anisotropic Raman spectrum of water (blue) together with two lineshape functions used to fit the two hydrogen-bond modes with the parameters listed in Tab.~\ref{tab:fit}. The experimental data have been taken from Ref.~\onlinecite{Castner1995}}
\label{fig:Raman1D}
\end{figure}

Fig.~\ref{fig:Raman1D} shows the 1D Raman spectrum of water in the relevant frequency range, which contains two spectroscopic features: the hydrogen-bond bending vibration around 50\,cm$^{-1}$ and the hydrogen-bond stretch vibration around 170\,cm$^{-1}$ (the librational mode around 600\,cm$^{-1}$ is not considered here since it is completely ouside our experimental observation window). Also shown in Fig.~\ref{fig:Raman1D} are fits to these two bands, assuming a linear response function
\R{
\begin{equation}
R_\mathrm{R}(\omega')=\Im \int_0^\infty \sin(\omega t)e^{-\frac{t}{2 T_1}}e^{-g(t)} e^{-i\omega' t} dt \label{eq:1D}
\end{equation}
The parameters are listed in Tab.~\ref{tab:fit} and will be justified later based on the fit of the echo observed in the 2D response.
}

\begin{figure}
\includegraphics[width=0.45\textwidth]{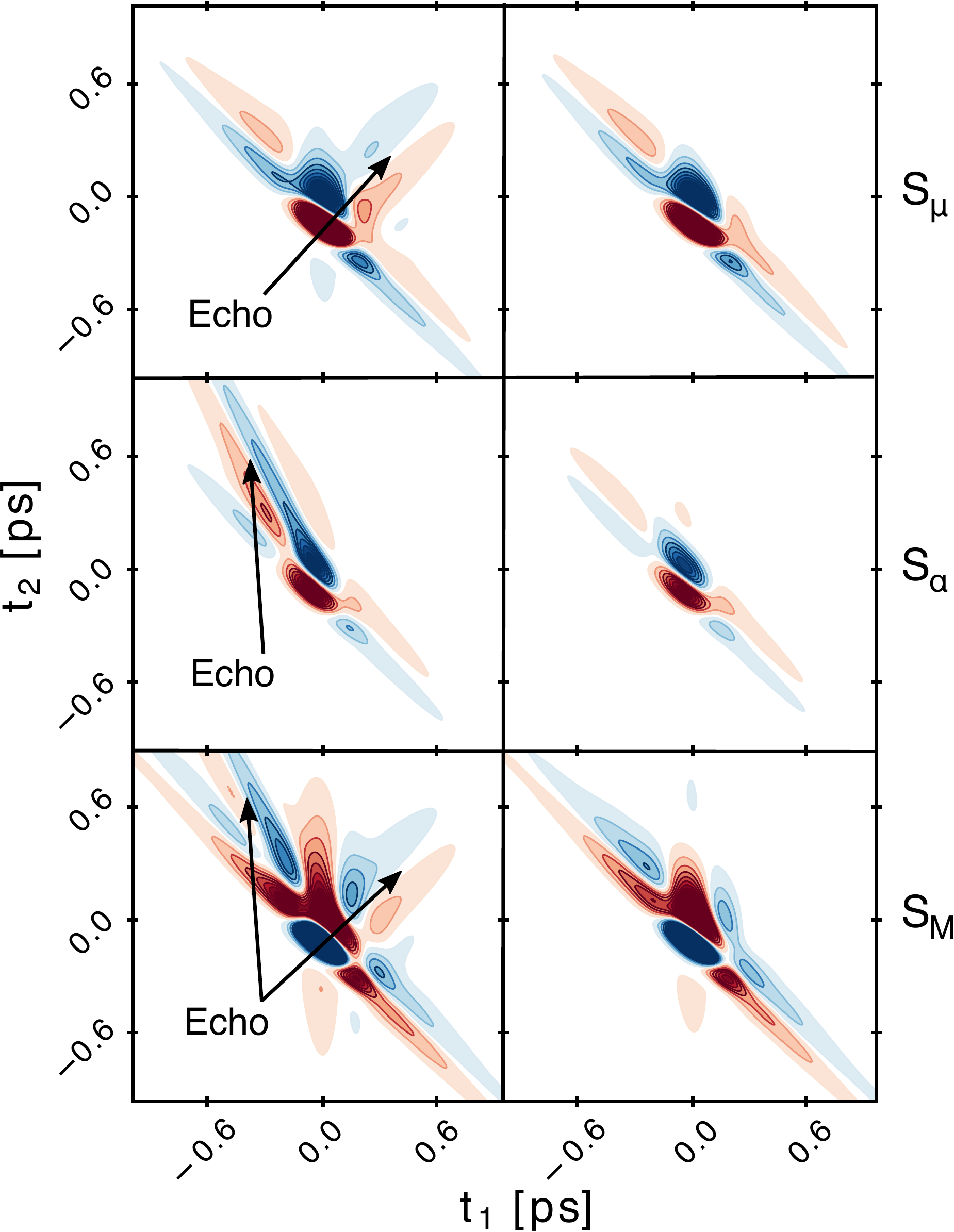}
\caption{Time domain signals of the responses $S^\mu$ (top), $S^\alpha$ (middle), and $S^\mathrm{M}$ (bottom) after convolution of Eq.~\ref{eq:paste} with the IRF Eq.~\ref{eq:convolution} and \ref{eq:IRF}. \R{The right column uses the parameters as in Tab.~\ref{tab:fit}, the left column sets $\tau_c=$100~ps and $T_1=1~$ps to more clearly demonstrate the echoes. Discernible echoes are marked by arrows.}} \label{fig:echo200}
\end{figure}

Starting with the  hydrogen-bond stretching vibration, Fig.~\ref{fig:echo200} (left column) shows the corresponding 2D results in the time domain, i.e., the convolution of Eq.~\ref{eq:paste} with the IRF. The lineshape parameters of \R{Tab.~\ref{tab:fit} were assumed, except for $\tau_c=$100~ps and $T_1=1~$ps, which in a first step was chosen unrealistically slow in order to more clearly demonstrate the resulting echoes (i.e., the same parameters as in Fig.~\ref{fig:echoes}).} Each one of the three types of anharmonicities give rise to a different 2D response. The echoes shown in Fig.~\ref{fig:echoes} survive the convolution and are marked by arrows in Fig.~\ref{fig:echo200}. In the case of a nonlinear dipole moment, the rephasing peak is in $R_\mathrm{RTT}^\mu$ and the echo is visible in the (upper right) RTT quadrant along $t_1 = t_2$. On the other hand, for $S_\alpha$ with the nonlinearity in the polarizability, the rephasing peak is in the $R_\mathrm{TRT}^\alpha$ response and the echo appears in the upper left quadrant along the $t_2 = -2 t_1$ due to the time transformation in Eq.~\ref{eq:paste}. Finally, $S_\mathrm{M}$ is a mix of both (albeit with different signs), since mechanical anharmonicity allows for all coherence pathways at the same time.  \R{Fig.~\ref{fig:echo200} (right column) shows the same, but now for $\tau_c=370$~fs and $T_1$=250~fs (which will be justified later based on a fit to the experimental data).The echoes are now masked by the instrument response function.}

\begin{figure}
\includegraphics[width=0.435\textwidth]{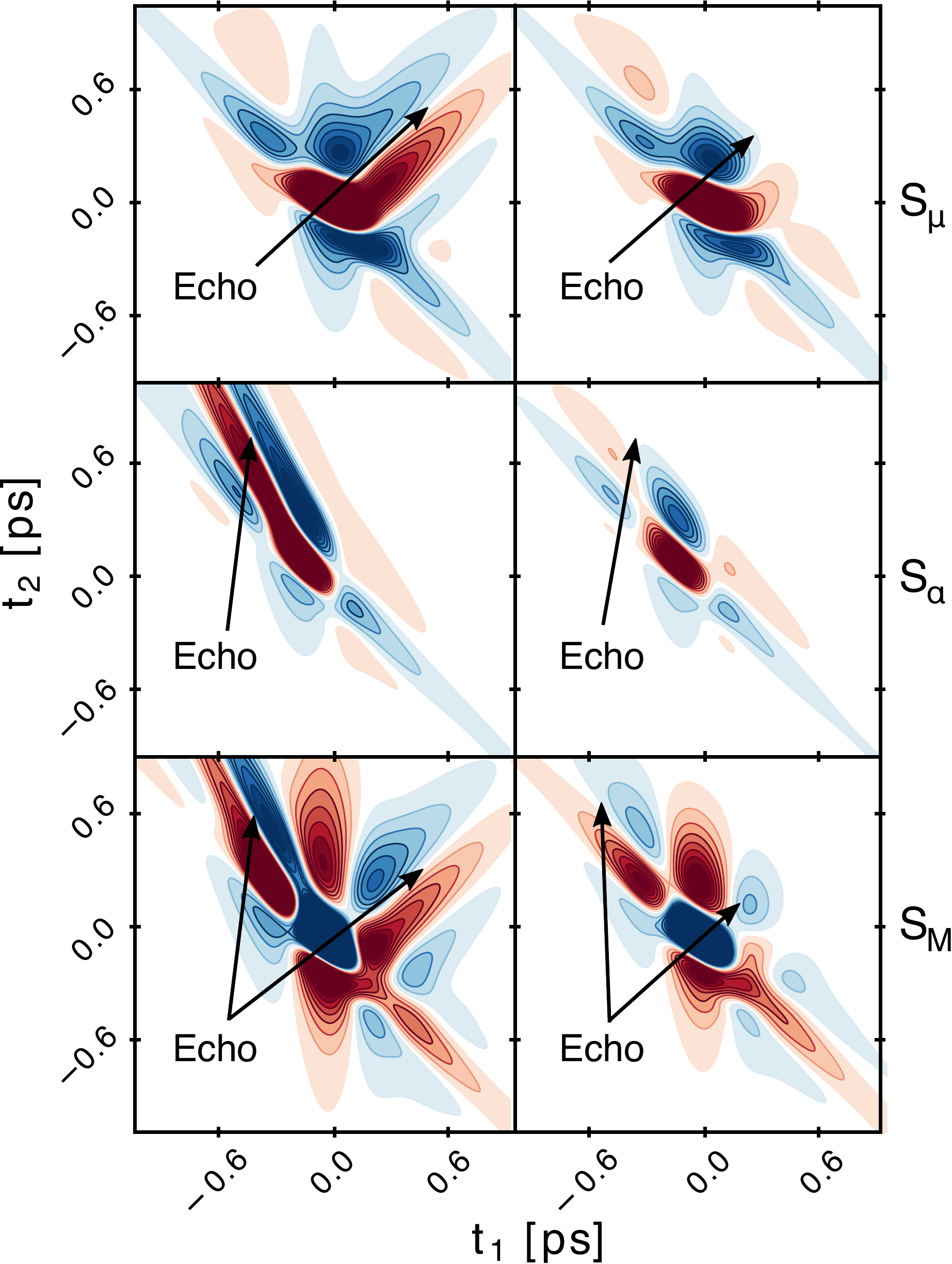}
\caption{Time domain signals of the responses $S^\mu$ (top), $S^\alpha$ (middle), and $S^\mathrm{M}$ (bottom) after convolution of Eq.~\ref{eq:paste} with the IRF Eqs.~\ref{eq:convolution} and \ref{eq:IRF}. \R{The right column uses the parameters as in Tab.~\ref{tab:fit}, the left column sets $\tau_c=$100~ps and $T_1=1~$ps to more clearly demonstrate the echoes. Discernible echoes are marked by arrows.}}
\label{fig:echo50}
\end{figure}

\R{Modelling the hydrogen-bond bending vibration around 50~cm$^{-1}$ on the same footing reveals similar results, see Fig.~\ref{fig:echo50}. The echoes are clearer in this case, even when using realistic parameters for $\tau_c$ and $T_1$ (Fig.~\ref{fig:echo50}, right column). In addition the signal intensity is significantly larger (Fig.~\ref{fig:echo200} has been upscaled by a factor 2), reflecting the the limited bandwidth of the THz pulses in the experiment, that peak at 50\,cm$^{-1}$ but only partially cover the 170\,cm$^{-1}$ band.\cite{Savolainen2013}}

\begin{figure*}
\centering
\includegraphics[width=0.8\textwidth]{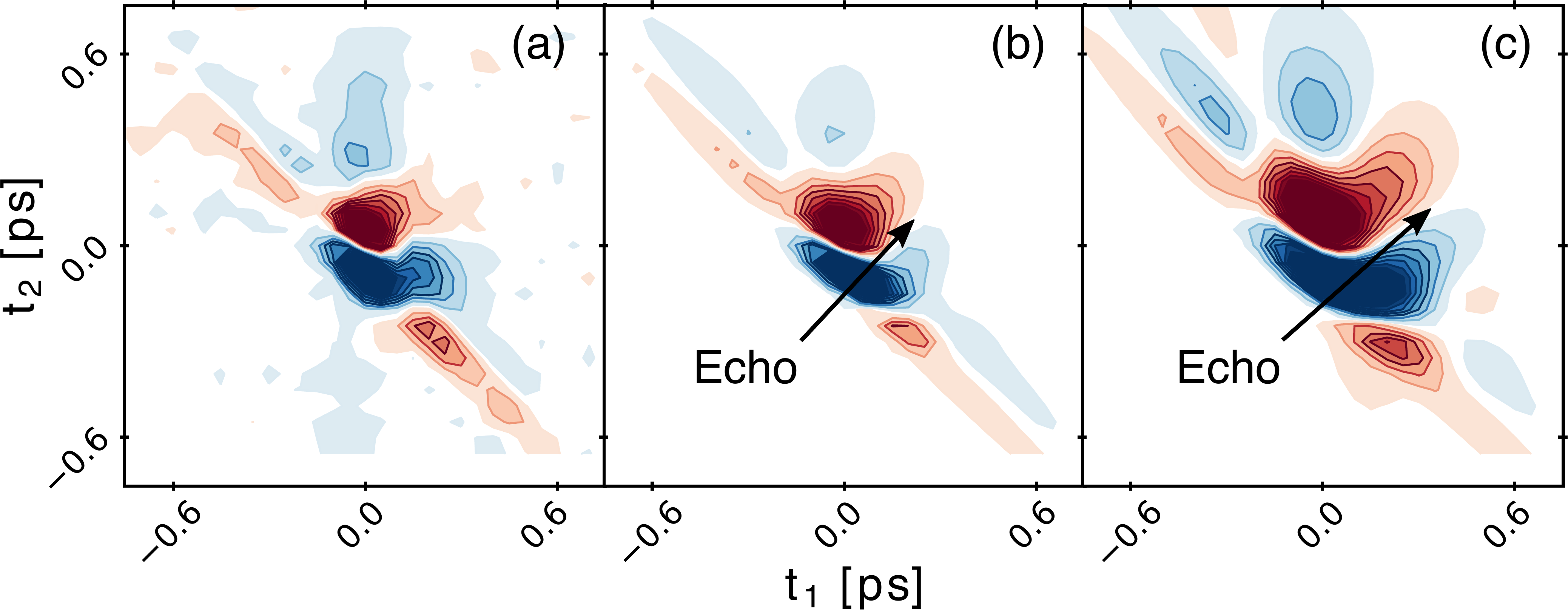}
\caption{ (a) Experimental water response at room temperature and signal calculated from the anharmonic oscillator model form (b) the hydrogen-bond stretching vibrations and (c) the hydrogen-bond bending vibrations with the parameters listed in Tab.~\ref{tab:fit}. Discernible echoes are marked by arrows. The experimental data have been taken from Ref.~\onlinecite{Savolainen2013}.}
\label{fig:experiment}
\end{figure*}

\begin{table}[b]
\centering
\caption{\R{Parameters obtained from the fit of the hydrogen-bond bending vibration and hydrogen-bond stretching vibration. } }
\label{tab:fit}
\begin{tabular}{l |  c c }
   &stretching & bending\\\hline
 $\omega$ [cm$^{-1}$]               & 165\footnotemark[1] & 45\footnotemark[1]\\
 $\Delta\omega$ [cm$^{-1}$]         & 75  & 40\\
 $\tau_c$ [fs]                      & 370 & 300\footnotemark[2] \\
 $T_1$ [fs]                         & 250 & 300\footnotemark[2] \\
 $\sigma_\mu/\sigma_\mathrm{M}$     & 1.4 & 0.96\\
 $\sigma_\alpha/\sigma_\mathrm{M}$  & -0.1 & 0.01\\
\end{tabular}
\footnotetext[2]{As a result of Eq.~\ref{eq:1D}, the peak position of the band does not coincide exactly with $\omega$}.
\footnotetext[2]{It has not been possible to minimize the RMSD for the hydrogen-bond bending vibration, since for any set of parameters, the RMSD is higher than that of the
instrument response function \textit{per se}. The optimization algorithm therefore converged to very short times $\tau_c$ and  $T_1$, in which case the molecular
response becomes $\delta$-shaped. In analogy to the hydrogen-bond bending vibration, we fixed $\tau_c=T_1=300$~fs, and then optimized all other parameters.}
\end{table}

\subsection{Fit of the Water Response}
\label{Sec:fit}
The free parameters of the model are $\sigma_\mu$, $\sigma_\alpha$, $\sigma_\mathrm{M}$, $T_1$, $\tau_c$, $\omega$ and $\Delta \omega$. In an iterative process, we varied the dephasing parameter $T_1$, $\tau_c$ to reproduce the 2D-response, and $\omega$ as well as $\Delta\omega$ to reproduce the position and width of the hydrogen-bond bending vibration in the 1D spectrum (Fig.~\ref{fig:Raman1D}). In addition, small variations of the delay-time zeros and the correction for the Gouy phase were allowed, which are not very accurately defined in the experiment.\cite{Savolainen2013} Once these nonlinear parameters are fixed, the 2D responses of Figs.~\ref{fig:echo200} and \ref{fig:echo50} (right column) can be considered a basis in which the experimental response is expanded in order to minimize the RMSD between experimental and fitted spectrum. Fig.~\ref{fig:experiment} shows that this procedure results in a remarkably good agreement with the experimental data, despite the simplicity of the model and the small number of parameters. Tab.~\ref{tab:fit} summarizes the resulting parameters for the two modes.

\R{The RMSD of the fit for the hydrogen-bond stretching vibration is smaller by a significant factor 0.7 as compared to that of the hydrogen-bond bending vibration. It has in fact not been possible to find a minimum in the RMSD for the hydrogen-bond bending vibration (see footnote in Tab.~\ref{tab:fit}), and in addition, the shift in $t_2$ required to obtain the best fit for hydrogen-bond bend vibration is larger (i.e., $\Delta t_2$=170~fs) than what we would think is its experimental uncertainty (the corresponding value for the hydrogen-bond stretching vibration is $\Delta t_2=35$~fs). We therefore suggest that the observed experimental signal originates predominantly from the hydrogen-bond stretching vibration, despite the fact its response is reduced to a certain extent as it is at the very edge of the experimentally accessible frequency window. We have not considered scenarios in which both bands contribute in parallel, or even more so, couple with each other, because the number of fitting parameters would be too large and the fitting problem would be under-determined.}

\begin{figure}
\centering
\includegraphics[width=0.45\textwidth]{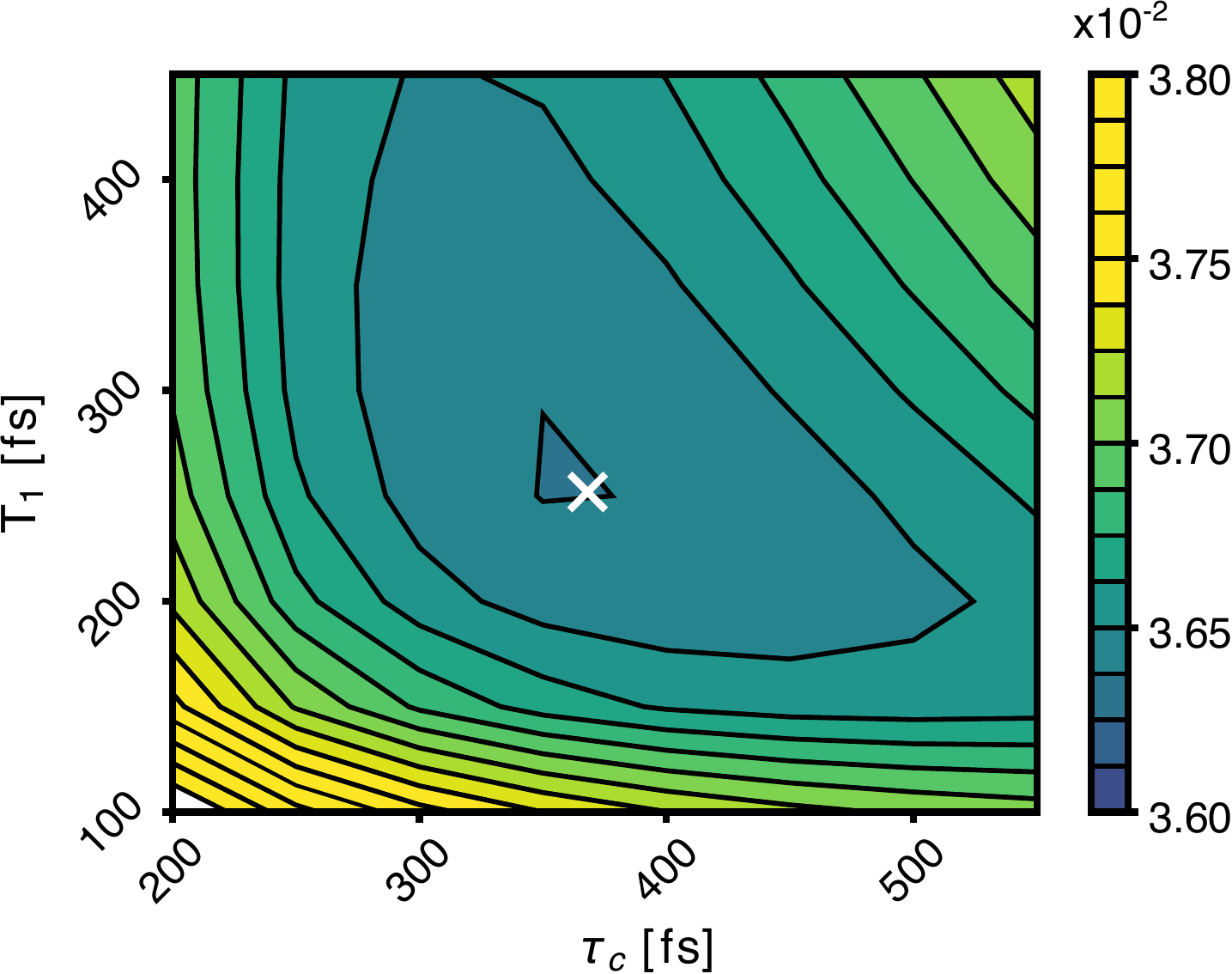}
\caption{\R{RMSD between experimental and simulated spectrum as a function of $\tau_c$ and $T_1$, considering the hydrogen-bond stretching vibration. The parameters for $\omega$ and $\Delta \omega$ were kept fixed to the values reported in Tab.~\ref{tab:fit}, since they are determined mostly by the 1D spectrum (Fig.~\ref{fig:Raman1D}), which in turn changes only very little when varying $\tau_c$ and $T_1$. The linear parameters $\sigma_\mu$, $\sigma_\alpha$, $\sigma_\mathrm{M}$, on the other hand, were optimized for each ($\tau_c$,$T_1$)-point in this plot. }}
\label{fig:rmsd}
\end{figure}

\begin{figure}
\centering
\includegraphics[width=0.35\textwidth]{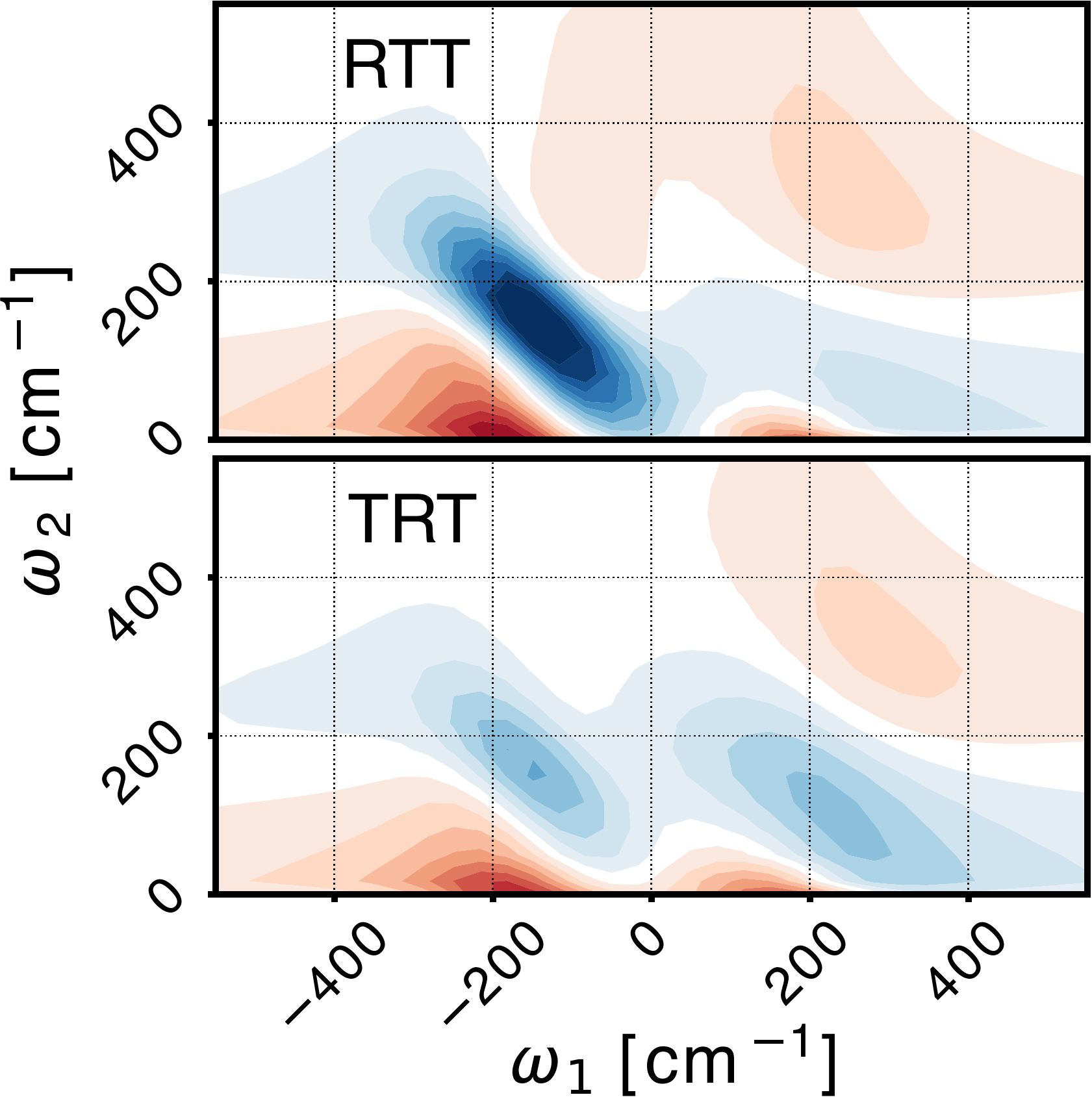}
\caption{\R{Same model as in Fig.~\ref{fig:experiment}b, but before convolution with the instrument response function and in the frequency domain, plotting $R_\mathrm{RTT}$ (top) and $R_\mathrm{TRT}$ (bottom) separately.}}
\label{fig:freqdomain}
\end{figure}

\R{Fig.~\ref{fig:rmsd} plots the RMSD of the fit for the hydrogen-bond stretching vibration as a function of $\tau_c$ and $T_1$, revealing that both parameters are not strongly correlated. As these two parameters determine the relative contribution of inhomogeneous vs homogeneous dephasing, we conclude that the signatures of the echo are still present in the 2D response. In that regard it is fortunate that the rephasing peak (a) in $R_\mathrm{RTT}^\mu$ (Fig.~\ref{fig:diag_thz}, bottom left) and $R_\mathrm{RTT}^\mathrm{M}$
(Fig.~\ref{fig:diag_mech}, bottom) have the same sign, while all other peaks have opposite sign (and the contribution of $R_\mathrm{RTT}^\mu$ is negligible, see Tab.~\ref{tab:fit}). Together with the corresponding weights of the two contributions ($\sigma_\mu/\sigma_\mathrm{M}$=1.4), the rephasing peak will actually dominate in the overall response. That is illustrated in Fig.~\ref{fig:freqdomain}, which shows the response in the  the frequency domain and without the convolution with the instrument response function for the same parameters as in in Fig.~\ref{fig:experiment}b; $R_\mathrm{RTT}$ is dominated by one rephasing peak that is strongly elongated along the diagonal due to the inhomogeneous broadening. As a result, an echo is discernible in in the time-domain data of Fig.~\ref{fig:experiment}b, despite the fact that is essentially completely masked in the individual contributions  of Fig.~\ref{fig:echo200} (right column).}

During the fit, the absolute intensity of the experimental signal was not considered. Likewise, the proportionality constants in Eq.~\ref{eq:linear}, $d\mu/dq$ and $d\alpha/dq$, are not known. Therefore, the three parameters $\sigma_\mu$, $\sigma_\alpha$, and $\sigma_\mathrm{M}$ are only known modulo an overall scaling factor, and only their relative contributions to the signal (e.g.., $\sigma_\mu/\sigma_\mathrm{M}$ and $\sigma_\alpha/\sigma_\mathrm{M}$) can be extracted. The sign of the overall signal can however be determined. The sign of the overall signal is given by the signs of $\sigma_\mathrm{M}$ times that of $d\alpha/dq$ (since $d\mu/dq$ enters quadratically in Eq.~\ref{eq:Rsep}). To reproduce the experimentally observed sign, we either have $\sigma_\mathrm{M}$ negative and $d\alpha/dq$ positive (which is what we assumed here), or \textit{vice versa}.

\section{Discussion and Conclusion}
We have shown that the experimental 2D-Raman-THz response of liquid water can be explained almost quantitatively by describing the hydrogen bond bending modes with the help of a very simple anharmonic oscillator model. \R{The model contains only six independent parameters, i.e., $\sigma_\mu/\sigma_\mathrm{M}$ and $\sigma_\alpha/\sigma_\mathrm{M}$ as well as $\omega_0$, $\Delta \omega$, $\tau_c$, and $T_1$. Despite the fact that the instrument response function masks the information content of the molecular 2D response to a significant extent, it is still sufficient to determine these parameters with confidence (Fig.~\ref{fig:rmsd}), if we assume the response originates from predominantly one of the two bands in the observation window of the experiment. We tentatively conclude that this dominating band is the hydrogen-bond stretching vibration at 170\,cm$^{-1}$, but we cannot exclude additional contributions from the  hydrogen-bond bending vibration around 50\,cm$^{-1}$, or from couplings between both bands. The librational mode around 600\,cm$^{-1}$, on the other hand, is completely outside our experimental observation window (Fig.~\ref{fig:irf}).}

The physical interpretation of $\sigma_\mu$, $\sigma_\alpha$, and $\sigma_\mathrm{M}$ is not straight forward, since the low frequency modes of water are collective in nature and presumably delocalized to a certain extent. \R{It is nonetheless worth noting that the biggest contribution to the signal originates from mechanical anharmonicity $\sigma_\mathrm{M}$ and the dipole nonlinearity $\sigma_\mu$, while the contribution from the nonlinearity in the polarizability is negligible with $\sigma_\alpha=0.1\sigma_\mathrm{M}$ (Tab.~\ref{tab:fit}). This observation is in stark contrast to results from recent MD work, which  revealed an echo for the TRT pulse sequence.\cite{Hamm2012a,Ito2015,Ikeda2015} That echo originated from the hindered rotation band around 600~cm$^{-1}$, which we however do not observe in the experiment due to the limited bandwidth. A large nonlinearity in the polarizability for that mode can be understood from the fact that a strict $\Delta J = 2$  selection rule would apply for a Raman interaction, together with a $\Delta J = 1$ selection rule for a THz interaction, in the limiting case of a free rather than hindered rotor.\cite{Hamm2017} This explanation obviously does not apply for the hydrogen-bond bending and stretching vibrations.}

\R{Regarding the line shape function, on the other hand, the parameters for $\tau_c$, $\Delta\omega$ and $T_1$ quantify the degree of inhomogeneous broadening of the hydrogen-bond-bending vibration (Tab.~\ref{tab:fit}). That is, with $\Delta\omega\tau_c\approx 5$, the lineshape function is in the ``slow-modulation'', or ``quasi-inhomogeneous'' limit.\cite{hamm11} At the same time, vibrational relaxation with $T_1=$250~fs contributes only 20\,cm$^{-1}$ to the total linewidth. The correlation time $\tau_c=$370~fs, in turn, is a measure of the lifetime of the hydrogen-bond networks giving rise to the hydrogen-bond stretching vibration. The typical lifetime of a single hydrogen bond is 1~ps,\cite{asbury04a,yere03,eaves05, perakis2011}, hence we conclude that
those modes are delocalized over $\approx3$ hydrogen bonds.}

The quality of the fit of Fig.~\ref{fig:experiment} is  much better than that of much more sophisticated calculations based on a water force field in connection with MD simulations.\cite{Hamm2014} The problem with the MD approach arises from the fact that a water force field needs to describe the thermodynamics and dynamics of water reasonably well in the first place, while the anharmonicities, that are the bottleneck of the 2D-Raman-THz signal, result from these constraints only in an indirect way. Anharmonicities are typically not fitted explicitly, except if very specific effects, such as nuclear quantum effects,\cite{Habershon2009} are to be described. Furthermore, in particular when polarizability is included in a water force field, the number of parameters is typically very large, and the problem is underdetermined. Electrical anharmonicity has a lot to do with the redistribution of charges during the motion of water molecules, and being able to quantify it might reveal guidelines to design better water models.\cite{Sidler2018} To that end, it would be important to infer the anharmonicity parameters $\sigma_\mu$, $\sigma_\alpha$, and $\sigma_\mathrm{M}$ from a MD simulation of a realistic water force field.

In conclusion, we think that 2D-Raman-THz spectroscopy has the highest information content as to date with regard to the intermolecular degrees of freedom water, but learning how to extract that information from the experimental response is challenging. The present work constitutes a significant step in that direction.\\

\textbf{Acknowledgments:} We thank Yoshitaka Tanimura for very valuable discussions. The work has been supported by the Swiss National Science Foundation (SNF) through the NCCR MUST.

\appendix
\section{Position Operator in an Anharmonic Eigenstate Basis} \label{app:anharmonic}

In the cubic anharmonic oscillator model, the harmonic oscillator Hamiltonian $\hat{H_0}$ is perturbed by $\sigma_M \hbar \omega q^3$.
\begin{equation}
\hat{H} = \hat{H_0} + \sigma_M \hbar \omega \hat q^3
\end{equation}
where $\hat q = \sqrt{m \omega / \hbar} \hat x$ is the unitless position operator, and $\sigma_M$ the size of the perturbation. To first order in $\sigma_M$, the eigenfunctions of $\hat{H}$ are expressed as linear combination of the harmonic eigenfunctions $| \varphi_n \rangle$. \cite{Claude1991}
\begin{align}
|  \varphi^{(anh)}_n \rangle &=| \varphi_n \rangle + \sigma_M \big( a_n | \varphi_{n+1} \rangle + b_n | \varphi_{n-1}\rangle \label{eq:eigenfun}  \\
& \quad  + c_n | \varphi_{n+3} \rangle + d_n | \varphi_{n-3} \rangle \big) \nonumber
\end{align}
with
\begin{align}
a_n &= -3 \left( \frac{n+1}{2} \right) ^{3/2}  \nonumber \\
b_n &= 3 \left( \frac{n}{2} \right) ^{3/2}  \nonumber \\
c_n &= -\frac{1}{3} \left[ \frac{(n+3)(n+2)(n+1)}{8} \right]^{1/2}  \nonumber \\
d_n &= \frac{1}{3} \left[ \frac{n(n-1)(n-2)}{8} \right] ^{1/2}.
\end{align}
Mechanical anharmonicity does not affect the position operator $\hat{q}$ \textit{per se}, but rather changes its matrix representation due to a change of the basis functions $\{| \varphi_i \rangle\} \rightarrow \{| \varphi^{(anh)}_i \rangle\}$. We will call $\hat{q}$ in a harmonic eigenfunction basis $\boldsymbol{q_H}$, and $\boldsymbol{q_A}$ if it is expressed in anharmonic  eigenfunctions. Expressing $(\boldsymbol{q_A})_{i,j}   \equiv \langle \varphi^{(anh)}_i | \hat{q} | \varphi^{(anh)}_j \rangle $ in terms of $(\boldsymbol{q_H})_{k,l} \equiv \langle \varphi_k | \hat{q} | \varphi_l \rangle$, using  Eq.~\ref{eq:eigenfun}, we get a large collection of terms:
\begin{align}
   (\boldsymbol{q_A})_{i,j}  =
(\boldsymbol{q_H})_{i,j} +  \sigma_M \big ( &a_j (\boldsymbol{q_H})_{i,j+1} + a_i (\boldsymbol{q_H})_{i+1,j} \nonumber \\
+ & b_j (\boldsymbol{q_H})_{i,j-1} + b_i (\boldsymbol{q_H})_{i-1,j} \nonumber \\
 + & c_j (\boldsymbol{q_H})_{i,j+3} + c_i (\boldsymbol{q_H})_{i+3,j}  \nonumber\\
  +& d_j (\boldsymbol{q_H})_{i,j-3} + d_i (\boldsymbol{q_H})_{i-3,j}  \big ),\label{eq:expansion}
\end{align}
where terms higher than first order in $\sigma_M$ have been discarded.
The matrix form of the position operator in harmonic oscillator eigenbasis is (see Eq.~\ref{eq:q}):
\begin{equation}
\boldsymbol{q_H} = \frac{1}{\sqrt{2}}
	\left(	\begin{matrix}
		0 & \sqrt{1} & 0 & 0 &  \\
		\sqrt{1} & 0 & \sqrt{2} & 0 & \\
		0 & \sqrt{2} & 0 &  \sqrt{3} &\\
		0 & 0  & \sqrt{3} & 0 & \ddots \\
		&  &  &  \ddots &\ddots
	\end{matrix} \right)	.
\end{equation}
A matrix with $(\boldsymbol{q_H})_{i,j+1}$ is the matrix $\boldsymbol{q_H}$ shifted up by one row:
\begin{equation}
\frac{1}{\sqrt{2}}
	\left(	\begin{matrix}
		\sqrt{1} & 0 & \sqrt{2} &  \\
		0 & \sqrt{2} & 0 &  \ddots\\
		0 & 0  & \sqrt{3} &  \ddots \\
		&  &  &  \ddots
	\end{matrix} \right)	.
\end{equation}
Hence, the elements $\sqrt{j} \delta_{i+1,j}$ are transformed to $\sqrt{i+1} \delta_{i,j}$ terms in that shifted matrix, and $\sqrt{i} \delta_{i-1, j}$ of $\boldsymbol{q_H}$ to $\sqrt{i} \delta_{i-2, j}$. Using this procedure, the shifted matrices in Eq.~\ref{eq:expansion} become:
\begin{align}
(\boldsymbol{q_H})_{i,j+1} &= \tfrac{1}{\sqrt{2}} \Big(  \sqrt{i+1} \delta_{i,j} + \sqrt{i} \delta_{i-2,j} \Big) \nonumber \\
(\boldsymbol{q_H})_{i+1,j} &= \tfrac{1}{\sqrt{2}} \Big( \sqrt{j+1}  \delta_{i,j} + \sqrt{j} \delta_{i,j-2} \Big)  \nonumber \\
(\boldsymbol{q_H})_{i,j-1} &= \tfrac{1}{\sqrt{2}} \Big( \sqrt{i} \delta_{i,j} + \sqrt{i+1} \delta_{i,j-2} \Big) \nonumber \\
(\boldsymbol{q_H})_{i-1,j} &= \tfrac{1}{\sqrt{2}} \Big( \sqrt{j} \delta_{i,j}  +  \sqrt{j+1} \delta_{i-2,j} \Big) \nonumber \\
(\boldsymbol{q_H})_{i,j+3} &= \tfrac{1}{\sqrt{2}} \Big( \sqrt{i+1} \delta_{i-2,j} +   \sqrt{i} \delta_{i-4,j} \Big) \nonumber \\
(\boldsymbol{q_H})_{i+3,j} &= \tfrac{1}{\sqrt{2}} \Big( \sqrt{j+1} \delta_{i,j-2} + \sqrt{j} \delta_{i,j-4} \Big) \nonumber \\
(\boldsymbol{q_H})_{i,j-3} &= \tfrac{1}{\sqrt{2}} \Big(  \sqrt{i} \delta_{i,j-2} + \sqrt{i+1} \delta_{i, j-4} \Big)  \nonumber \\
(\boldsymbol{q_H})_{i-3,j} & = \tfrac{1}{\sqrt{2}} \Big(   \sqrt{j} \delta_{i-2,j} + \sqrt{j+1} \delta_{i-4, j} \Big)
\end{align}
which include zero- ($\delta_{i,j}$), two- ($\delta_{i-2,j}$ and $\delta_{i,j-2}$), and four-quantum ($\delta_{i-4,j}$ and $\delta_{i,j-4}$) contributions. Collecting all terms, we obtain:

\begin{figure*}
	\centering
	\includegraphics[width=.8\textwidth]{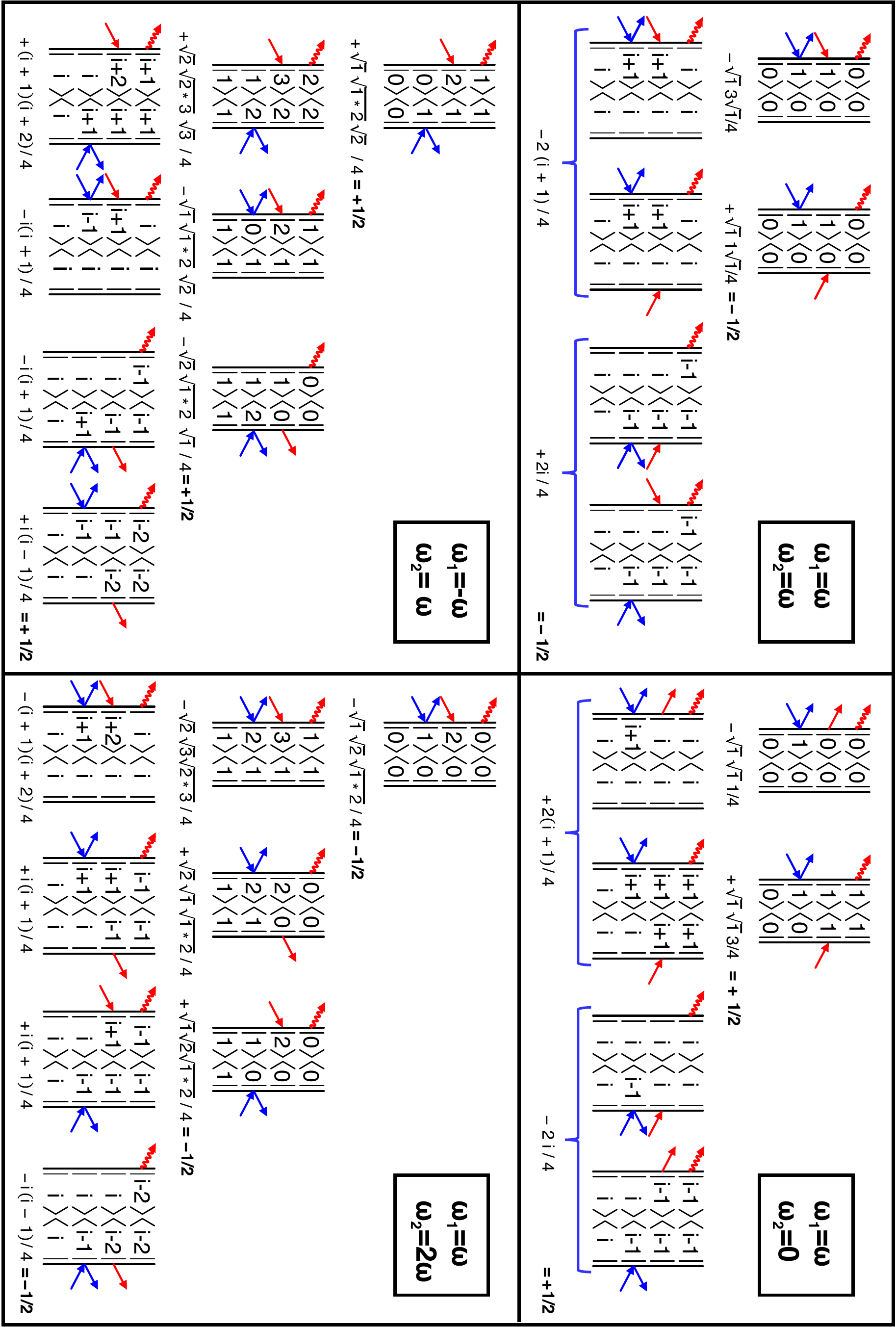}
	\caption{Collection of all Feynman diagrams responsible for the four peaks  ($\omega, \omega$), ($\omega, 0$), ($-\omega, \omega$),  and  ($\omega, 2\omega$) present in $R_\mathrm{RTT}^\mu$. The pathways are ordered by their initial state, and the intensity of each pathway is denoted below the corresponding Feynman diagrams.}
	\label{fig:pathways}
\end{figure*}

\begin{align}
\tfrac{1}{\sqrt{2}} \Big( a_j \sqrt{i+1} + a_i \sqrt{j+1} + b_j \sqrt{i} + b_i \sqrt{j} \Big) \delta_{i,j} & = \nonumber\\
- \tfrac{3}{2} \big( 2i + 1 \big) \delta_{i,j} \nonumber\\
\tfrac{1}{\sqrt{2}} \Big( a_j \sqrt{i} + b_i \sqrt{j+1} + c_j \sqrt{i+1} + d_i \sqrt{j} \Big) \delta_{i-2,j}& =\nonumber\\
\tfrac{1}{2} \sqrt{i}\sqrt{i-1} \delta_{i-2,j}\nonumber\\
\tfrac{1}{\sqrt{2}} \Big( a_i \sqrt{j} + b_j \sqrt{i+1} + c_i \sqrt{j+1} + d_j \sqrt{i} \Big) \delta_{i,j-2} &= \nonumber \\
\tfrac{1}{2} \sqrt{j}\sqrt{j-1} \delta_{i,j-2} \nonumber\\
\tfrac{1}{\sqrt{2}} \Big( c_j \sqrt{i}  + d_i \sqrt{j+1} \Big) \delta_{i-4,j} &=0 \\
\label{eq:fourQ}
\end{align}
We see that the $\delta_{i-4,j}$ elements cancel (likewise for $\delta_{i,j-4}$), and only zero- and two-quantum transitions remain in $\boldsymbol{q_A}$, in addition to the one quantum transitions from $\boldsymbol{q_H}$.
Inserting Eq.~\ref{eq:fourQ} into Eq.~\ref{eq:expansion} we get
\begin{align}
\boldsymbol{q_A} = \boldsymbol{q_H} + \sigma_M \boldsymbol{q_M},
\end{align}
where the second term describes the correction to the position matrix due to anharmonicity with the final expression of $\boldsymbol{q_M}$ given in Eq.~\ref{eq:qM}.

\section{Temperature Independence of the Response Function} \label{sec:temperature}

Temperature enters via the thermal population of the initial density matrix $\rho_\mathrm{eq}$. In IR spectroscopy, one usually assumes that only the ground state is initially populated due to $\hbar \omega \gg k_B T$, but that assumption no longer holds in THz spectroscopy, and the contributions from many initial states have to be considered. Fig.~\ref{fig:pathways} exemplifies that for $R_\mathrm{RTT}^\mu$ by showing all contributing Feynman diagrams sorted according to their initial state. Since that response functions contain all four nonzero peaks (a), (b), (c), and (d), the result is universal and applies in the same way for the other response functions. It can be seen that the overall intensities do not depend on initial state and hence also not on temperature. However, a prerequisite for this to work is the assumption \R{that the lineshape functions depend on energy level differences only (Eq.~\ref{eq_vibRelax})}, and that the energy spacing between states is equidistant, i.e. that of a harmonic oscillator, and that the effect of mechanical anharmonicity only enters via the softened selection rules that allow for zero- and two-quantum transitions. \R{Fig.~\ref{fig:diag_mech} demonstrates that this is a good approximation (see discussion in the main text).}


\begin{thebibliography}{10}

\bibitem{Bertie1996}
J.~E. Bertie and Z.~Lan,
\newblock {Infrared Intensities of Liquids XX: The Intensity of the OH
  Stretching Band of Liquid Water Revisited, and the Best Current Values of the
  Optical Constants of H2O(l) at 25$^\circ$C between 15,000 and 1 cm$^{-1}$},
\newblock Appl. Spectrosc. {\bf 50}, 1047 (1996).

\bibitem{tan93}
Y.~Tanimura and S.~Mukamel,
\newblock {2-Dimensional femtosecond vibrational spectroscopy of liquids},
\newblock J. Chem. Phys. {\bf 99}, 9496 (1993).

\bibitem{okumura97a}
K.~Okumura and Y.~Tanimura,
\newblock {Femtosecond two-dimensional spectroscopy from anharmonic vibrational
  modes of molecules in the condensed phase},
\newblock J. Chem. Phys. {\bf 107}, 2267 (1997).

\bibitem{VBout1991}
D.~{Vanden Bout}, L.~J. Muller, and M.~Berg,
\newblock {Ultrafast Raman Echoes in Liquid Acetonitrile},
\newblock Phys. Rev. Lett. {\bf 67}, 3700 (1991).

\bibitem{Inaba1993}
R.~Inaba, K.~Tominaga, M.~Tasumi, K.~A. Nelson, and K.~Yoshihara,
\newblock Observation of homogeneous vibrational dephasing in benzonitrile by
  ultrafast raman echoes,
\newblock Chem. Phys. Lett. {\bf 211}, 183  (1993).

\bibitem{Muller1993}
L.~J. Muller, D.~{Vanden Bout}, and M.~Berg,
\newblock {Broadening of Vibrational Lines by Attractive Forces: Ultrafast
  Raman Echo Experiments in a CH3I:CDCl3 Mixture},
\newblock J. Chem. Phys. {\bf 99}, 810 (1993).

\bibitem{Tokmakoff1997}
A.~Tokmakoff, M.~J. Lang, D.~S. Larsen, G.~R. Fleming, V.~Chernyak, and
  S.~Mukamel,
\newblock {Two-Dimensional Raman Spectroscopy of Vibrational Interactions in
  Liquids},
\newblock Phys. Rev. Lett. {\bf 79}, 2702 (1997).

\bibitem{Blank1999}
D.~A. Blank, L.~J. Kaufman, and G.~R. Fleming,
\newblock {Fifth-Order Two-Dimensional Raman Spectra of CS2 are Dominated by
  Third-Order Cascades},
\newblock J. Chem. Phys. {\bf 111}, 3105 (1999).

\bibitem{Blank2000}
D.~A. Blank, L.~J. Kaufman, and G.~R. Fleming,
\newblock {Direct Fifth-Order Electronically Nonresonant Raman Scattering from
  CS2 at Room Temperature},
\newblock J. Chem. Phys. {\bf 113}, 771 (2000).

\bibitem{Kaufman2002}
L.~J. Kaufman, J.~Heo, L.~D. Ziegler, and G.~R. Fleming,
\newblock {Heterodyne-Detected Fifth-Order Nonresonant Raman Scattering from
  Room Temperature CS2},
\newblock Phys. Rev. Lett. {\bf 88}, 207402 (2002).

\bibitem{Golonzka2000}
O.~Golonzka, N.~Demird{\"{o}}ven, M.~Khalil, and A.~Tokmakoff,
\newblock {Separation of Cascaded and Direct Fifth-Order Raman Signals using
  Phase-Sensitive Intrinsic Heterodyne Detection},
\newblock J. Chem. Phys. {\bf 113}, 9893 (2000).

\bibitem{Kubarych2003}
K.~J. Kubarych, C.~J. Milne, and R.~J. Miller,
\newblock {Fifth-Order Two-Dimensional Raman Spectroscopy: A New Direct Probe
  of the Liquid State},
\newblock Int. Rev. Phys. Chem. {\bf 22}, 497 (2003).

\bibitem{Li2008}
Y.~L. Li, L.~Huang, R.~J. Miller, T.~Hasegawa, and Y.~Tanimura,
\newblock {Two-Dimensional Fifth-Order Raman Spectroscopy of Liquid Formamide:
  Experiment and Theory},
\newblock J. Chem. Phys. {\bf 128}, 234507 (2008).

\bibitem{Kuehn2009}
W.~Kuehn, K.~Reimann, M.~Woerner, and T.~Elsaesser,
\newblock {Phase-Resolved Two-Dimensional Spectroscopy based on Collinear
  n-Wave Mixing in the Ultrafast Time Domain},
\newblock J. Chem. Phys. {\bf 130}, 164503 (2009).

\bibitem{Kuehn2011}
W.~Kuehn, K.~Reimann, M.~Woerner, T.~Elsaesser, R.~Hey, and U.~Schade,
\newblock {Strong Correlation of Electronic and Lattice Excitations in
  GaAs/AlGaAs Semiconductor Quantum Wells Revealed by Two-Dimensional Terahertz
  Spectroscopy},
\newblock Phys. Rev. Lett. {\bf 107}, 067401 (2011).

\bibitem{Kuehn2011a}
W.~Kuehn, K.~Reimann, M.~Woerner, T.~Elsaesser, and R.~Hey,
\newblock {Two-Dimensional Terahertz Correlation Spectra of Electronic
  Excitations in Semiconductor Quantum Wells},
\newblock J. Phys. Chem. B {\bf 115}, 5448 (2011).

\bibitem{Elsaesser2015}
T.~Elsaesser, K.~Reimann, and M.~Woerner,
\newblock {Focus: Phase-resolved Nonlinear Terahertz Spectroscopy - From Charge
  Dynamics in Solids to Molecular Excitations in Liquids},
\newblock J. Chem. Phys. {\bf 142}, 212301 (2015).

\bibitem{Lu2016}
J.~Lu, Y.~Zhang, H.~Y. Hwang, B.~K. Ofori-Okai, S.~Fleischer, and K.~A. Nelson,
\newblock {Nonlinear Two-Dimensional Terahertz Photon Echo and Rotational
  Spectroscopy in the Gas Phase},
\newblock Proc. Natl. Acad. Sci. USA {\bf 113}, 11800 (2016).

\bibitem{Somma2016}
C.~Somma, G.~Folpini, K.~Reimann, M.~Woerner, and T.~Elsaesser,
\newblock {Phase-Resolved Two-Dimensional Terahertz Spectroscopy including
  Off-Resonant Interactions beyond the $\chi$(3)limit},
\newblock J. Chem. Phys. {\bf 144}, 184202 (2016).

\bibitem{Finneran2016}
I.~A. Finneran, R.~Welsch, M.~A. Allodi, T.~F. Miller, and G.~A. Blake,
\newblock {Coherent Two-Dimensional Terahertz-Terahertz-Raman Spectroscopy},
\newblock Proc. Natl. Acad. Sci. USA {\bf 113}, 6857 (2016).

\bibitem{Finneran2017}
I.~A. Finneran, R.~Welsch, M.~A. Allodi, T.~F. Miller, and G.~A. Blake,
\newblock {2D THz-THz-Raman Photon-Echo Spectroscopy of Molecular Vibrations in
  Liquid Bromoform},
\newblock J. Phys. Chem. Lett. {\bf 8}, 4640 (2017).

\bibitem{Savolainen2013}
J.~Savolainen, S.~Ahmed, and P.~Hamm,
\newblock {Two-Dimensional Raman-THz Spectroscopy of Water},
\newblock Proc. Natl. Acad. Sci. USA {\bf 110}, 20402 (2013).

\bibitem{Shalit2017}
A.~Shalit, S.~Ahmed, J.~Savolainen, and P.~Hamm,
\newblock {Terahertz Echoes Reveal the Inhomogeneity of Aqueous Salt
  Solutions},
\newblock Nat. Chem. {\bf 9}, 273 (2017).

\bibitem{Berger2018}
A.~Berger, G.~Ciardi, P.~Hamm, and A.~Shalit,
\newblock {The Impact of Nuclear Quantum Effects on the Structural
  Inhomogeneity of Liquid Water},
\newblock (submitted).

\bibitem{hamm11}
P.~Hamm and M.~T. Zanni,
\newblock {\em {Concepts and Methods of 2D Infrared Spectroscopy}},
\newblock Cambridge University Press, Cambridge, 2011.

\bibitem{Cho1999}
M.~Cho,
\newblock {Theoretical Description of the Vibrational Echo Spectroscopy by
  Time-Resolved Infrared-Infrared-Visible Difference-Frequency Generation},
\newblock J. Chem. Phys. {\bf 111}, 10587 (1999).

\bibitem{Hamm2012}
P.~Hamm and J.~Savolainen,
\newblock {Two-Dimensional-Raman-Terahertz Spectroscopy of Water: Theory},
\newblock J. Chem. Phys. {\bf 136}, 094516 (2012).

\bibitem{Hamm2012a}
P.~Hamm, J.~Savolainen, J.~Ono, and Y.~Tanimura,
\newblock {Note: Inverted Time-Ordering in Two-Dimensional-Raman-Terahertz
  Spectroscopy of Water},
\newblock J. Chem. Phys. {\bf 136}, 1 (2012).

\bibitem{Hamm2014}
P.~Hamm,
\newblock {2D-Raman-THz Spectroscopy: A Sensitive Test of Polarizable Water
  Models},
\newblock J. Chem. Phys. {\bf 141} (2014).

\bibitem{Hamm2017}
P.~Hamm and A.~Shalit,
\newblock {Perspective: Echoes in 2D-Raman-THz Spectroscopy},
\newblock J. Chem. Phys. {\bf 146} (2017).

\bibitem{Steffen1996}
T.~Steffen, J.~T. Fourkas, and K.~Duppen,
\newblock {Time resolved four- and six-wave mixing in liquids. I. Theory},
\newblock Journal of Chemical Physics {\bf 105}, 7364 (1996).

\bibitem{Steffen1998}
T.~Steffen and K.~Duppen,
\newblock {Population relaxation and non-Markovian frequency fluctuations in
  third- and fifth-order Raman scattering},
\newblock Chemical Physics {\bf 233}, 267 (1998).

\bibitem{saito98}
S.~Saito and I.~Ohmine,
\newblock {Off-resonant fifth-order nonlinear resonse of water and
  {\{}CS{\$}{\_}2{\$}{\}}: Analysis based on normal modes},
\newblock J. Chem. Phys. {\bf 108}, 240 (1998).

\bibitem{ma00}
A.~Ma and R.~M. Stratt,
\newblock {Fifth-order {\{}R{\}}aman spectrum of an atomic liquid: simulation
  and instantaneous normal mode calculation},
\newblock Phys. Rev. Lett. {\bf 85}, 1004 (2000).

\bibitem{Jansen2000}
T.~l.~C. Jansen, J.~G. Snijders, and K.~Duppen,
\newblock {The third- and fifth-order nonlinear Raman response of liquid CS2
  calculated using a finite field nonequilibrium molecular dynamics method},
\newblock The Journal of Chemical Physics {\bf 113}, 307 (2000).

\bibitem{Okumura2003}
K.~Okumura and Y.~Tanimura,
\newblock {Energy-level diagrams and their contribution to fifth-order Raman
  and second-order infrared responses: Distinction between relaxation models by
  two-dimensional spectroscopy},
\newblock Journal of Physical Chemistry A {\bf 107}, 8092 (2003).

\bibitem{saito06}
S.~Saito and I.~Ohmine,
\newblock {Fifth-order two-dimensional Raman spectroscopy of liquid water,
  crystalline ice Ih and amorphous ices: Sensitivity to anharmonic dynamics and
  local hydrogen bond network structure},
\newblock J. Chem. Phys. {\bf 125}, 84506 (2006).

\bibitem{Hasegawa2006}
T.~Hasegawa and Y.~Tanimura,
\newblock {Calculating Fifth-Order Raman Signals for various Molecular Liquids
  by Equilibrium and Nonequilibrium Hybrid Molecular Dynamics Simulation
  Algorithms},
\newblock J. Chem. Phys. {\bf 125} (2006).

\bibitem{Cho2000}
M.~Cho,
\newblock {Theoretical Description of Two-Dimensional Vibrational Spectroscopy
  by Infrared-Infrared-Visible Sum Frequency Generation},
\newblock Phys. Rev. A. {\bf 61}, 12 (2000).

\bibitem{Zhao2000}
W.~Zhao and J.~C. Wright,
\newblock {Doubly Vibrationally Enhanced Four Wave Mixing: The Optical Analog
  to 2D NMR},
\newblock Phys. Rev. Lett. {\bf 84}, 1411 (2000).

\bibitem{Guo2009}
R.~Guo, F.~Fournier, P.~M. Donaldson, E.~M. Gardner, I.~R. Gould, and D.~R.
  Klug,
\newblock {Detection of Complex Formation and Determination of Intermolecular
  Geometry through Electrical Anharmonic Coupling of Molecular Vibrations using
  Electron-Vibration-Vibration Two-Dimensional Infrared Spectroscopy},
\newblock Phys. Chem. Chem. Phys {\bf 11}, 8417 (2009).

\bibitem{Grechko2018}
M.~Grechko, T.~Hasegawa, F.~D'Angelo, H.~Ito, D.~Turchinovich, Y.~Nagata, and
  M.~Bonn,
\newblock {Coupling Between Intra-and Intermolecular Motions in Liquid Water
  Revealed by Two-Dimensional Terahertz-Infrared-Visible Spectroscopy},
\newblock Nat. Commun. {\bf 9} (2018).

\bibitem{Ito2014}
H.~Ito, T.~Hasegawa, and Y.~Tanimura,
\newblock {Calculating Two-Dimensional THz-Raman-THz and Raman-THz-THz Signals
  for Various Molecular Liquids: The Samplers},
\newblock J. Chem. Phys. {\bf 141}, 124503 (2014).

\bibitem{Ito2015}
H.~Ito, J.~Y. Jo, and Y.~Tanimura,
\newblock {Notes on Simulating Two-Dimensional Raman and Terahertz-Raman
  Signals with a Full Molecular Dynamics Simulation Approach},
\newblock Struct. Dyn. {\bf 2} (2015).

\bibitem{Ikeda2015}
T.~Ikeda, H.~Ito, and Y.~Tanimura,
\newblock {Analysis of 2D THz-Raman Spectroscopy Using a non-Markovian Brownian
  Oscillator Model with Nonlinear System-Bath Interactions},
\newblock J. Chem. Phys. {\bf 142} (2015).

\bibitem{Pan2015}
Z.~Pan, T.~Wu, T.~Jin, Y.~Liu, Y.~Nagata, R.~Zhang, and W.~Zhuang,
\newblock {Low Frequency 2D Raman-THz Spectroscopy of Ionic Solution: A
  Simulation Study},
\newblock J. Chem. Phys. {\bf 142} (2015).

\bibitem{Ito2016}
H.~Ito, T.~Hasegawa, and Y.~Tanimura,
\newblock {Effects of Intermolecular Charge Transfer in Liquid Water on Raman
  Spectra},
\newblock J. Phys. Chem. Lett. {\bf 7}, 4147 (2016).

\bibitem{Ito2016b}
H.~Ito and Y.~Tanimura,
\newblock {Simulating Two-Dimensional Infrared-Raman and Raman Spectroscopies
  for Intermolecular and Intramolecular Modes of Liquid Water},
\newblock J. Chem. Phys. {\bf 144} (2016).

\bibitem{Atkins2010}
P.~W. Atkins and R.~S. Friedman,
\newblock {\em Molecular Quantum Mechanics},
\newblock Oxford University Press, 2010.

\bibitem{Claude1991}
C.~Cohen-Tannoudji, B.~Diu, and F.~Laloe,
\newblock {\em Quantum Mechanics, Volume 2},
\newblock Wiley-VCH, 1991.

\bibitem{Ahmed2014}
S.~Ahmed, J.~Savolainen, and P.~Hamm,
\newblock {The Effect of the Gouy Phase in Optical-Pump-THz-Probe
  Spectroscopy},
\newblock Opt. Express {\bf 22}, 4256 (2014).

\bibitem{Castner1995}
E.~W. Castner, Y.~J. Chang, Y.~C. Chu, and G.~E. Walrafen,
\newblock {The Intermolecular Dynamics of Liquid Water},
\newblock J. Chem. Phys. {\bf 102}, 653 (1995).

\bibitem{asbury04a}
J.~B. Asbury, T.~Steinel, K.~Kwak, S.~A. Corcelli, C.~P. Lawrence, J.~L.
  Skinner, and M.~D. Fayer,
\newblock {Dynamics of water probed with vibrational echo correlation
  spectroscopy},
\newblock J. Chem. Phys. {\bf 121}, 12431 (2004).

\bibitem{yere03}
S.~Yeremenko, M.~S. Pshenichnikov, and D.~A. Wiersma,
\newblock {Hydrogen-bond dynamics in water explored by heterodyne-detected
  photon echo},
\newblock Chem. Phys. Lett. {\bf 369}, 107 (2003).

\bibitem{eaves05}
J.~D. Eaves, J.~J. Loparo, C.~J. Fecko, S.~T. Roberts, A.~Tokmakoff, and P.~L.
  Geissler,
\newblock {Hydrogen bonds in liquid water are broken only fleetingly},
\newblock Proc. Natl. Acad. Sci. USA {\bf 102}, 13019 (2005).

\bibitem{perakis2011}
F.~Perakis, S.~Widmer, and P.~Hamm,
\newblock {Two-dimensional infrared spectroscopy of isotope-diluted ice Ih},
\newblock Journal of Chemical Physics {\bf 134}, 1 (2011).

\bibitem{Habershon2009}
S.~Habershon, T.~E. Markland, and D.~E. Manolopoulos,
\newblock {Competing quantum effects in the dynamics of a flexible water
  model},
\newblock J. Chem. Phys. {\bf 131}, 024501 (2009).

\bibitem{Sidler2018}
D.~Sidler, M.~Meuwly, and P.~Hamm,
\newblock {An efficient water force field calibrated against intermolecular THz
  and Raman spectra},
\newblock J. Chem. Phys. {\bf 148}, 244504 (2018).

\end{thebibliography}

\end{document}